\documentclass[a4paper,10pt]{article}
\pdfoutput=1 

\usepackage{jheppub} 

\usepackage{color,graphicx,slashed,multirow,amssymb,amsmath,multirow}
\usepackage{braket}
\usepackage{graphicx}
\usepackage[utf8]{inputenc}
\usepackage[section]{placeins}
\usepackage{xcolor,colortbl}

\usepackage{subcaption}
\usepackage{ragged2e} 

\makeatletter
\renewcommand{\fnum@figure}{\textbf{Fig.~\thefigure}}
\renewcommand{\fnum@table}{\textbf{Tab.~\thetable}}
\long\def\@makecaption#1#2{%
  \vskip\abovecaptionskip
  \setlength{\@tempdima}{\linewidth}%
  \par\noindent
  \begin{minipage}{\@tempdima}%
    \small\textbf{#1.}\ \justifying #2%
  \end{minipage}\par
  \vskip\belowcaptionskip}
\makeatother

\usepackage{hyperref}
\usepackage{cleveref}
\usepackage[normalem]{ulem}

\usepackage[shortlabels]{enumitem}
\AtBeginDocument{\usepackage{booktabs}}

\renewcommand{\arraystretch}{1.2}

\newcommand{\Lag}{{\mathcal L}}

\newcommand{\SUD}{{SU(2)_{\rm D}}}

\newcommand{\Z}{{\mathbb Z}}

\newcommand{\mg}{{\sc\small MG5\_aMC}}

\newcommand{\pythia}{{\sc\small Pythia8.2}}

\newcommand{\be}{\begin{equation}}
\newcommand{\ee}{\end{equation}}
\def\bsp#1\esp{\begin{split}#1\end{split}}
\def\bpm{\begin{pmatrix}}
\def\epm{\end{pmatrix}}

\newcommand{\com}[1]{\iffalse #1 \fi}

\usepackage{xcolor}

\usepackage[normalem]{ulem}

\renewcommand\figurename{Fig.}
\renewcommand\tablename{Tab.}
\crefname{equation}{Eq.}{Eqs.}
\crefname{figure}{Fig.}{Figs.}
\crefname{tabular}{Tab.}{Tabs.}

\title{\boldmath The Muonic Portal to Vector Dark Matter: \\connecting precision muon physics, cosmology, and colliders}

\author[a,b]{Alexander Belyaev}

\author[c,d]{, Luca Panizzi}

\author[e,a]{, Nakorn Thongyoi}

\author[f]{, Franz Wilhelm}

\affiliation[a]{School of Physics and Astronomy, University of Southampton, Highfield, Southampton SO17 1BJ, UK}
\affiliation[b]{Particle Physics Department, Rutherford Appleton Laboratory, Chilton, Didcot, Oxon OX11 0QX, UK}
\affiliation[c]{Dipartimento di Fisica, Università della Calabria, Arcavacata di Rende, I-87036, Cosenza, Italy}
\affiliation[d]{INFN, Gruppo Collegato di Cosenza, Arcavacata di Rende, I-87036, Cosenza, Italy}
\affiliation[e]{Khon Kaen Particle Physics and Cosmology Theory Group (KKPaCT), Department of Physics, Faculty of Science, Khon Kaen University, 123 Mitraphap Rd, Khon Kaen 40002, Thailand}
\affiliation[f]{Department of Physics and Astronomy, Uppsala University, Box 516, SE-751 20 Uppsala, Sweden}

\emailAdd{a.belyaev@soton.ac.uk}
\emailAdd{luca.panizzi@unical.it}
\emailAdd{nakorn.thongyoi@gmail.com}
\emailAdd{franzwilhelm42@gmail.com}

\abstract{
We present a comprehensive study of the Muonic Portal to Vector Dark Matter (MPVDM), a minimal extension of the Standard Model featuring a new $SU(2)_D$ gauge symmetry and vector-like muons that mediate interactions between the dark sector and the muon sector.
We show that the MPVDM can simultaneously reproduce the observed dark matter relic abundance and accommodate scenarios consistent with the current experimental determination of the muon anomalous magnetic moment, $(g-2)_\mu$, as well as scenarios allowing for a non-zero new physics contribution to
$(g-2)_\mu$.
One of the key results of this work is the identification of a {\it generic off-resonance velocity-suppression mechanism} that allows light ($\lesssim 1$~GeV) vector dark matter to evade stringent CMB constraints near $2m_{\mathrm{DM}}\simeq m_{H_D}$.
A five-dimensional parameter scan combining cosmological, collider, and precision constraints shows that scenarios admitting a non-zero contribution to $(g-2)_\mu$ favour sub-GeV dark matter realised near the scalar resonance with a dark gauge coupling $g_D\!\sim\!10^{-3}$ and TeV-scale vector-like muons, while scenarios consistent with a Standard-Model-like $(g-2)_\mu$ allow a broad viable dark matter mass range from sub-GeV to multi-TeV.
By recasting ATLAS and CMS searches for $\mu^+\mu^-$ final states with missing transverse energy, we derive a lower bound of approximately 850~GeV on the vector-like muon masses.
We further identify distinctive multi-lepton collider signatures, including six-, eight-, and ten-muon final states as well as mixed muon--electron topologies with displaced electron pairs, providing striking and well-motivated targets for searches at the LHC and future colliders.
}

\begin{document}

\maketitle

\section{Introduction}
\label{sec:intro}

Understanding the nature of Dark Matter (DM) remains one of the greatest puzzles in modern particle physics and cosmology. While overwhelming observational evidence, spanning galactic to cosmological scales, supports the existence of DM, decades of experimental efforts have only confirmed its gravitational interactions. Dedicated observations of the Cosmic Microwave Background (CMB) anisotropies by the Planck experiment imply that the amount of DM is approximately five times greater than that of baryonic matter~\cite{Planck:2018vyg}. However, key questions about DM, such as its spin, mass, non-gravitational interactions, stability mechanism, number of associated states, and the potential mediators of interactions with Standard Model (SM) particles, remain unanswered. The evidence for DM provides, arguably, the strongest experimental indication of Beyond the Standard Model (BSM) physics.

A further long-discussed potential tension between Standard Model (SM) predictions and observed data concerns the anomalous magnetic moment of the muon~\cite{Aoyama:2020ynm}, $a_\mu \equiv (g-2)_\mu/2$, which remains a topic of intense interest in particle physics due to its potential to reveal the nature of physics beyond the SM.

The precise measurements of $a_\mu$ from Fermilab (FNAL)~\cite{Muong-2:2025xyk}, combined with earlier measurements from Brookhaven (BNL)~\cite{Muong-2:2002wip,Muong-2:2004fok,Muong-2:2006rrc}, have improved the accuracy of $a_\mu$ by about a factor of two, offering a sharper probe of possible deviations from the SM. The latest combined world average is
\begin{align}
    a_\mu^{\rm EXP} = 116592071.5(14.5)\times10^{-11},
    \label{eq:g-2 exp}
\end{align}
with a precision of 124 parts per billion. Further improvements are expected from the ongoing analysis of the Fermilab Run~6 data and from the forthcoming J-PARC experiment~\cite{MIBE2011242,Abe:2019thb}, which employs an independent measurement technique.

While the experimental determinations are now in strong mutual agreement, the theoretical situation remains complex. The 2020 Muon $g-2$ Theory Initiative white paper~\cite{Aoyama:2020ynm} gave the benchmark SM prediction
\begin{align}
    a_\mu^{\rm SM,WP2020} = 116591810(43)\times10^{-11},
    \label{eq:g-2 SM old}
\end{align}
which was consistent with previous theoretical evaluations~\cite{PhysRevLett.109.111808,Aoyama:2019ryr,PhysRevD.67.073006,PhysRevD.88.053005,Davier:2017zfy,Hoferichter:2019mqg,Davier:2019can,KURZ2014144,PhysRevD.70.113006,PhysRevD.95.054026,Colangelo:2017fiz,Hoferichter:2018kwz,Gerardin:2019vio,BIJNENS2019134994,Colangelo:2019uex,PhysRevLett.124.132002,COLANGELO201490}. At that time, comparing Eq.~\eqref{eq:g-2 SM old} with the then-current experimental value yielded
\begin{align}
    \Delta a_\mu^{\rm EXP,t} \equiv a_\mu^{\rm EXP} - a_\mu^{\rm SM,WP2020} = 249(48)\times10^{-11},
    \label{eq:delta_amu_exp_t}
\end{align}
corresponding to a tension of roughly $5\sigma$. Updating to the 2025 experimental average while keeping the 2020 theory baseline gives
\begin{align}
    \Delta a_\mu = 261(45)\times10^{-11},
\end{align}
which slightly increases the central discrepancy (though this comparison mixes results from different epochs and should be treated only illustratively).

The largest source of theoretical uncertainty arises from the hadronic vacuum polarization (HVP) contribution. Two complementary approaches exist. The first is the data-driven dispersive method, which uses experimental measurements of $e^+e^- \to \text{hadrons}$ to determine the HVP via a dispersion relation~\cite{Hoferichter:2019mqg,Davier:2019can,Keshavarzi:2019abf,Colangelo_2019}. The second is a first-principles computation via lattice QCD~\cite{Budapest-Marseille-Wuppertal:2017okr,RBC:2018dos,FermilabLattice:2019ugu,Giusti:2019xct,Gerardin:2019rua}. Historically, the dispersive method had smaller quoted uncertainties, while lattice results were limited by systematics.

This situation changed when the BMW collaboration reported in 2020~\cite{Borsanyi:2020mff} a sub-percent lattice determination of the HVP, obtaining a larger $a_\mu^{\rm SM}$ and thereby reducing the experiment--theory discrepancy to about $1.5\sigma$. That result, however, generated debate because it was in tension with global electroweak fits via its effect on $\Delta\alpha_{\rm had}^{(5)}(m_Z)$~\cite{Crivellin:2020zul}. More recently, the Fermilab/HPQCD/MILC collaboration published a new high-precision lattice computation of the total HVP contribution~\cite{Bazavov:2024PRL}, finding a result consistent with the BMW value within uncertainties. The agreement between these two independent lattice collaborations strengthens the reliability of the lattice-based HVP evaluation, which yields a higher SM prediction and consequently a smaller difference with experiment.

Meanwhile, the experimental picture evolved as the CMD-3 collaboration reported new cross-section measurements for $e^+e^- \to \text{hadrons}$~\cite{CMD-3:2023alj}, which are systematically higher than earlier results from CMD-2~\cite{CMD-2:2003gqi,CMD-2:2005mvb,Aulchenko:2006dxz,CMD-2:2006gxt}, SND~\cite{Achasov:2006vp}, KLOE~\cite{KLOE:2008fmq,KLOE:2010qei,KLOE:2012anl,KLOE-2:2017fda}, BaBar~\cite{BaBar:2012bdw}, and BESIII~\cite{BESIII:2015equ}. If correct, CMD-3’s higher hadronic cross sections would increase the dispersive HVP estimate, raising the SM value of $a_\mu$ and largely resolving the previous tension. However, this dataset remains in tension with the older, highly consistent measurements.

In response to these developments, the Muon $g-2$ Theory Initiative released an updated white paper in 2025~\cite{Aliberti:2025WP}. This new evaluation incorporates the CMD-3 and BESIII data, the latest lattice results, and refined estimates of the hadronic light-by-light contribution. Adopting the lattice-informed HVP inputs in this update results in a major upward shift of the total SM prediction, which now reads
\begin{align}
    a_\mu^{\rm SM,WP2025} = 116592033(62)\times10^{-11},
    \label{eq:g-2 SM new}
\end{align}
corresponding to a total theoretical uncertainty of 530~ppb. When compared with the new experimental world average of Eq.~\eqref{eq:g-2 exp}, the difference becomes
\begin{align}
 \Delta a_\mu^{\rm EXP,c}  \equiv  a_\mu^{\rm EXP} - a_\mu^{\rm SM,WP2025} = 38(63)\times10^{-11},
 \label{eq:delta_amu_exp_c}
\end{align}
indicating that, within current uncertainties, there is no statistically significant tension between experiment and the SM prediction.

Thus, the field presently entertains two scientifically distinct possibilities. 
In the first, referred to as the \textit{tension scenario}, one retains the pre-2023 dispersive-based evaluations, for which the excess 
$\Delta a_\mu^{\rm EXP,t} \approx (2.5\pm0.5)\times10^{-9}$ (approximately $5\sigma$) persists, 
motivating models with new particles coupled to muons and photons. 
In the second, termed the \textit{compatibility scenario}, following the 2025 Theory Initiative update, 
the SM and experiment are consistent within uncertainties, albeit relying on the CMD-3 cross-sections, 
which are higher than all previous experiments.

In what follows, we therefore consider both cases: (i) regions of parameter space capable of explaining 
the positive excess $\Delta a_\mu^{\rm EXP,t} \sim 5\sigma$, including theoretical and experimental uncertainties, 
and (ii) regions compatible with a nearly null deviation according to the 2025 theory evaluation. 
This dual approach ensures that our model remains consistent with the evolving theoretical and experimental 
picture of the muon anomalous magnetic moment.


The $a_\mu$ anomaly and the DM problem provide strong motivation for exploring BSM physics. Currently, several models aim to simultaneously address these issues, including the Scalar Dark Matter (SDM) models~\cite{Davoudiasl:2012ig,Hong:2016uou,NA64:2021xzo,Das:2021zea,Qi:2021rhh}, the SDM with Vector-Like Leptons (VLLs)~\cite{PhysRevD.88.056017,PhysRevLett.127.061802,Barducci:2018esg,Kopp_2014,bai2021muon,Arora:2022uof,Kawamura:2022uft}, the Vector Dark Matter (VDM) with scalar portal~\cite{Ghorbani:2021yiw,Yang:2018fje}, and the VDM with VLL portal~\cite{Chowdhury:2021tnm}. Related non-Abelian vector dark matter scenarios have also been studied in the literature, including multicomponent dark sectors~\cite{Elahi:2019jeo} and models in which heavy fermions induce electromagnetic multipole interactions of complex vector dark matter~\cite{Hisano:2020qkq,Chu:2023zbo}.

In our previous papers~\cite{Belyaev:2022shr,Belyaev:2022zjx}, we proposed a novel theoretical construct in which the DM is a gauge vector from a non-Abelian group, interacting with the fermionic sector of the SM via the mediation of new vector-like (VL) partners of SM fermions, which we termed the \emph{Fermionic Portal to Vector Dark Matter} (FPVDM). We comprehensively studied scenarios where the VL fermion doublet is composed of VL top quarks, examining both their cosmological and collider phenomenology. In contrast to the multipole-based scenarios of Refs.~\cite{Hisano:2020qkq,Chu:2023zbo}, the defining feature of the FPVDM framework is a renormalisable fermionic portal connecting the dark sector directly to SM fermions, which in the present work is specialised to the muon sector.

In this paper, we investigate the possibility of simultaneously addressing the DM problem and the muon anomalous magnetic moment by focusing on a specific realization of this framework, where the portal is mediated by a doublet of muon partners:\footnote{For clarity, we use different notations for the new fields depending on the basis. The interaction eigenstates are denoted by $\psi_{\mu_D}$ and $\psi_{\mu^\prime}$, while the mass eigenstates are denoted by $\mu_D$ and $\mu^\prime$.}
\[
\Psi = \begin{pmatrix} \mu_D \\ \mu'  \end{pmatrix}\;.
\]
We abbreviate this realization as MPVDM from now on. The theoretical foundation of the MPVDM model is based on the FPVDM framework. The connection between the SM and dark sectors is established via Yukawa interactions involving both SM and VL muons. In addition to SM particles in the loop, $a_\mu$ receives contributions from new gauge bosons, scalars, and fermions in the MPVDM, which we have thoroughly evaluated and collectively refer to as \emph{New Physics} (NP) contributions. Using these results, we demonstrate that the MPVDM model has the potential to address both the $a_\mu $ anomaly and the DM problem simultaneously, leading to a specific new scenario with distinctive signatures.

This paper is organised as follows. 
In~\Cref{sec:model} we summarise the theoretical structure of the MPVDM framework. 
In~\Cref{sec:g-2} we review the relevant contributions to $a_\mu$ and identify the regions of parameter space compatible with current measurements. 
\Cref{sec:cosmo} examines the dark-matter candidate and the parameter space that survives cosmological constraints, including relic density, direct detection, and indirect detection bounds. 
In~\Cref{sec:collider pheno} we present collider limits derived from pair production of vector-like muon partners, leading to the characteristic $\mu^+\mu^- + \slashed{E}_T$ signature. 
We also outline a set of novel multi-lepton final states that arise uniquely in this model and constitute particularly striking LHC targets. 
In~\Cref{sec:combined} we analyse the interplay between the $a_\mu$, collider, and dark-matter constraints, isolating the region in which all bounds are simultaneously satisfied. 
\Cref{sec:benchmarks} summarises and discusses in detail the representative benchmark points that encapsulate the phenomenology of the model in both the short- and long-lived regimes.
Finally, we summarise our findings in~\Cref{sec:conclusions} and discuss future directions for exploring the MPVDM framework.

\section{The MPVDM model}
\label{sec:model}

In this section we briefly review the main features of the FPVDM model and its realisation in the muon sector, but we refer to our previous studies~\cite{Belyaev:2022shr,Belyaev:2022zjx} for more details. 

The FPVDM extends minimally a class of models where the DM candidates are massive gauge bosons associated with a non-Abelian symmetry group, $SU(2)_D$. These gauge bosons acquire mass through a spontaneous symmetry breaking mechanism in the dark sector, mediated by a scalar doublet $\Phi_D$. In this minimal construction, the Higgs portal is the sole interaction channel between the dark and SM sectors, with DM stability ensured by custodial symmetry within the scalar sector~\cite{Hambye:2008bq}. 
In FPVDM scenarios, on the other hand, the quartic interaction in the scalar sector is assumed to be negligibly small, or even absent at all. The interaction between the dark sector and the SM occurs via new fermions that transform non-trivially under $SU(2)_D \times U(1)_Y$, and interact with $\Phi_D$ and right-handed SM fermions sharing the same hypercharge through a Yukawa coupling.
The DM stability is ensured by imposing a global $U(1)_D$ symmetry under which only the new fields transform non-trivially. This is necessary because otherwise, due to the pseudo-real nature of the fundamental representation of $SU(2)$, a further Yukawa interaction involving $\Phi_D^c$ would be possible, which would break $SU(2)_D$ to nothing after $\Phi_D$ acquires a vacuum expectation value.
Instead, in the FPVDM the symmetry breaking pattern is $SU(2)_D\times U(1)_D \to U(1)_D^d$, where $U(1)_D^d$ is associated to the diagonal generator of the broken group.
With the $U(1)_D$ phase assignments $Y_D=\frac{1}{2}$ for dark scalar and fermion doublets, and $Y_D=0$ for vector triplet, there is still an invariance under the subgroup $\Z_2 \equiv (-1)^{Q_D}$, where $Q_D=T^3_D+Y_D$.
The quantum numbers of SM and new particles under the full gauge group of the theory is summarised in \Cref{tab:particlesQN}.

\setlength{\tabcolsep}{3pt}
\setlength{\arraycolsep}{0pt}
\begin{table}[htbp]
\centering
\begin{tabular}{c|cc|c||c|r}
\toprule
 & $SU(2)_L$ & $U(1)_Y$ & $\SUD$ & $\Z_2$ & $Q_D$\\
\midrule
$\Phi_{D}=\left(\begin{array}{c} \varphi^0_{D+ \frac{1}{2} } \\ \varphi^0_{D-\frac{1}{2} } \end{array}\right)$ & $\mathbf{1}$ & $0$ & $\mathbf{2}$ 
& $\begin{array}{c} - \\ + \end{array}$ 
& $\begin{array}{r} +1 \\ 0 \end{array}$ 
\\[2pt]
\midrule
\multirow{2}{*}{$\Psi=\left(\begin{array}{c} \psi_{D} \\ \psi \end{array}\right)$} & \multirow{2}{*}{$\mathbf{1}$} & \multirow{2}{*}{$Q$} & \multirow{2}{*}{$\mathbf{2}$} & $-$ & $+1$\\
& & & & $+$ & $0$\\[2pt]
\midrule
$V^D_{\mu}=\left(\begin{array}{c} V^0_{D+\mu} \\ V^0_{D0\mu} \\ V^0_{D-\mu} \end{array}\right)$ & $\mathbf{1}$ & $0$ & $\mathbf{3}$ 
& $\begin{array}{c} - \\ + \\ - \end{array}$ 
& $\begin{array}{r} +1 \\ 0 \\ -1 \end{array}$ 
\\[2pt]
\bottomrule
\end{tabular}
\caption{\label{tab:particlesQN}The quantum numbers  of the new particles under the Electro-Weak (EW) and $\SUD$ gauge groups. In this table the flavour of the SM fermion(s) which the new fermionic fields interact with is not specified, so the notation of the fermion doublet is left generic.}
\end{table}

The most general Lagrangian for this scenario takes the following form:
\begin{eqnarray}
\Lag &\supset& - \frac{1}{4} (V_{\mu\nu}^i)^2|_{B,W^i,V^i_D} + \bar{f}^{\rm SM} i \slashed{D} f^{\rm SM} + \bar{\Psi} i \slashed{D} \Psi + | D_\mu \Phi_H |^2 + | D_\mu \Phi_D |^2 - V(\Phi_H, \Phi_D) \nonumber \\
&-& (y \bar{f}^{\rm SM}_L \Phi_H f^{\rm SM}_R + y^\prime \bar{\Psi}_L \Phi_D f^{\rm SM}_R + h.c.) - M_\Psi \bar{\Psi} \Psi \;,
\label{eq: L of D sector}
\end{eqnarray}
where $V(\Phi_H, \Phi_D)$ is the scalar potential and is given by
\begin{equation}
 V(\Phi_H,\Phi_D) = - \mu^2 \Phi_H^\dagger \Phi_H - \mu_D^2 \Phi_D^\dagger \Phi_D + \lambda (\Phi_H^\dagger \Phi_H)^2 + \lambda_D (\Phi_D^\dagger \Phi_D)^2 + \lambda_{HD} (\Phi_H^\dagger \Phi_H)(\Phi_D^\dagger \Phi_D)\;.
 \label{eq:scalarpotential}
\end{equation}

Though the potential for the Higgs portal \( \Phi_H \) and \( \Phi_D \) could play a vital role in the phase transition, baryogenesis, and gravitational waves, it is not directly related to the current subject of this study. Therefore, for simplicity, we consider a minimal scenario where the quartic coupling \( \lambda_{HD} \) is negligibly small. Moreover, the smallness of this portal guarantees that the SM-like Higgs observables at the LHC remain intact.\\

Specific realisations of the model can be obtained by assigning the hypercharge of $\Psi$ and selecting which fermion(s) of the SM are allowed to interact with it. In this paper, we aim to explore the potential of FPVDM to explain the DM observables and $a_\mu$ values for both scenarios: (i) the case of an experimental excess over the SM prediction~\cite{Aoyama:2020ynm}, as indicated by \Cref{eq:delta_amu_exp_t}, and (ii) the case in which such an excess is absent. 
To address both scenarios (i) and (ii), we introduce a dark fermionic VL doublet that couples to the SM muon, 
$\Psi = (\psi_{\mu_D}, \psi_{\mu^\prime})^T$.
After EW and dark symmetry breaking, all the new states acquire masses. The new Yukawa term allows a mixing only between the SM and $\psi_{\mu^\prime}$ interaction eigenstates, leading to the mass eigenstates $\mu$ and $\mu^\prime$. On the other hand, we can identify the interaction ($\psi_{\mu_D}$) and mass ($\mu_D$) eigenstates for the $\Z_2$-odd state. The masses in the muon sector are thus given by 
\begin{equation}
 m_{\mu_D}=M_\Psi
 \quad\text{and}\quad
 m_{\mu,\mu^\prime}^2=\frac{1}{4} \left[y^2 v^2 + y^{\prime2} v_D^2 + 2 M_\Psi^2 \mp \sqrt{(y^2 v^2 + y^{\prime2} v_D^2 + 2 M_\Psi^2)^2-8y^2v^2M_\Psi^2}\right]
 \;,
 \label{eq: fermionic masses}
\end{equation}
where $v$ and $v_D$ are the vacuum expectations values (VEVs) of the Higgs and $\Phi_D$ doublets respectively.
The mixing angles between left-handed and right-handed chiral projections of SM and dark fermions are given by 
\begin{equation}
\sin\theta_{R} = \sqrt{\frac{m_{\mu^\prime}^2 - m_{\mu_D}^2}{m_{\mu^\prime}^2 - m_\mu^2}} \quad\text{and}\quad \sin\theta_{L} = \frac{m_\mu}{m_{\mu_D}}\sin\theta_{R}\;.
\end{equation}

A mass hierarchy in the fermionic sector of the MPVDM emerges from \Cref{eq: fermionic masses}:
\begin{align}
    m_\mu<m_{\mu_D}\leq m_{\mu^\prime}.
    \label{eq:fermionic mass hierarchy}
\end{align}
The two Yukawa couplings $y$ and $y^\prime$ can be determined as functions of the masses and VEVs as:
\begin{equation}
\label{eq:yukawas}
y = \sqrt{2} \frac{m_\mu m_{\mu^\prime}}{m_{\mu_D} v},\quad y^\prime = \sqrt2 \frac{\sqrt{(m_{\mu^\prime}^2 - m_{\mu_D}^2)(m_{\mu_D}^2 - m_\mu^2)}}{m_{\mu_D} v_D}\;.
\end{equation}
The new fermion sector is completely decoupled in the limit $m_{\mu^\prime}=m_{\mu_D}$, for which $y=y_{\rm SM}=\sqrt2 \frac{m_\mu}{v}$, $y^\prime=0$, $\sin\theta_{L}=\sin\theta_{R}=0$, and the pure SM scenario is restored.\\

In the scalar sector, the two $\mathbb{Z}_2$-even physical degrees of freedom that remain after symmetry breaking in the SM and dark sectors can mix to form the mass eigenstates $H$ (the SM Higgs boson) and $H_D$ (the Higgs boson from $SU(2)_D$)\footnote{Note that $H_D$ does not carry a dark charge; we simply use this notation to clearly distinguish this Higgs boson from the SM one, $H$}, with masses:
\begin{equation}
m_{H,H_D}^2 = \lambda v^2 + \lambda_D v_D^2 \mp \sqrt{(\lambda v^2 - \lambda_D v_D^2)^2 + \lambda_{HD}^2 v^2 v_D^2}\;.
\end{equation}
The mixing angle in the scalar sector is defined in terms of the scalar masses as:
\begin{equation}
\sin\theta_S = \sqrt{2 \frac{m_{H_D}^2 v^2 \lambda - m_H^2 v_D^2 \lambda_D}{m_{H_D}^4 - m_H^4}}\;.
\end{equation}
This mixing angle is a free parameter of the theory, which we will set to zero in the following, corresponding to setting $\lambda_{HD} = 0$ in \Cref{eq:scalarpotential}. Even in the absence of explicit tree-level mixing induced by the quadratic term, the scalars can still mix at one-loop via their interactions with fermions. The consequences of this loop-induced mixing, which can also affect Higgs-related observables, are beyond the scope of this analysis and will be addressed in future work.

In the gauge boson sector, the tree-level masses of the $SU(2)_D$ vectors read
\begin{equation}
m_V\equiv m_{V^\prime} = m_{V_D} = g_D\frac{v_D}{2} \label{eq:VPmass} \;,
\end{equation}
where $V_D\equiv V^0_{D\pm}$ and $V^\prime\equiv V^0_{D0}$, following the notation of~\cite{Belyaev:2022shr,Belyaev:2022zjx}. 
At loop level, the mass degeneracy is broken by the kinetic mixing of $\gamma$-$Z$-$V^\prime$ states, and by the different corrections to the masses of $V^\prime$ and $V_D$ due to the different particles circulating in the loops. The one-loop mass splitting between $V_D$ and $V^\prime$ is given by
\begin{align}
m_{V_D}-m_{V^\prime}=\frac{1}{3}\frac{g_D^2m_{\mu^\prime}^2}{32 \pi^2 m_{V_D}}\left(\frac{m_{\mu^\prime}^2-m_{\mu_D}^2}{m_{\mu^\prime}^2}\right)^2\;.
\label{eq:simple_mass_splitting_2}
\end{align}
The derivation of the analogous expression can be found in~\cite{Belyaev:2022shr}, where it was obtained for the top-quark portal. In the present muon-portal case, the mass splitting differs by an overall factor of $1/3$, which arises from the absence of the quark colour factor.
This radiative mass splitting between the $V_D$ and $V^{\prime}$ bosons plays a very important role in the determination of DM relic density and DM indirect detection rates. \\

From the determination of the spectrum of the model, it is now clear that the only possible DM candidate of the model is the $\Z_2$-odd gauge boson $V_D$, as it is the only particle which is electrically (and colour) neutral. This also implies that consistency with the cosmological observations requires $m_{\mu_D}>m_{V_D}$. In the following, $m_{\mathrm{DM}}$ and $m_{V_D}$ are used interchangeably, depending on the context, with $m_{\mathrm{DM}} \equiv m_{V_D}$.

The number of independent parameters of the model is fixed by the experimental observables, including the masses of the 
muon and Higgs boson. In general, there are six free parameters
for MPVDM:
$g_D,m_{V_D}, m_{H_D}, m_{\mu^\prime}, m_{\mu_D},$ $ \sin\theta_S$.
However, since we focus only on the effects of the fermionic portal, we neglect the quartic term in the scalar potential at tree level by setting $\sin\theta_S =0$. This makes the parameter space under study  five-parametric:
\begin{equation}
g_D,m_{V_D}, m_{H_D}, m_{\mu_D} \text{ and } m_{\mu^\prime}. 
\label{eq:pars}
\end{equation}
The dependent quantities can be written in terms of these parameters as follows:
\begin{equation}
\begin{array}{c}
v = 2 \frac{m_W}{g}\ \ , \ \ 
v_D = 2 \frac{m_{V_D}}{g_D}\ \ , \ \ 
\lambda = \frac{g^2m_H^2}{8m_W^2}\ \ , \ \ 
\lambda_D = \frac{g_D^2m_{H_D}^2}{8m_{V_D}^2}\ \ , \ \
\lambda_{HD}=0\\
y = \frac{g\;m_\mu}{\sqrt{2} m_W}\frac{m_{\mu^\prime}}{m_{\mu_D}}
\ \ , \ \
y^\prime =\frac{g_D\sqrt{(m_{\mu^\prime}^2 - m_{\mu_D}^2)(m_{\mu_D}^2 - m_\mu^2)}}{\sqrt{2} m_{\mu_D} m_{V_D}}.
\end{array}
\end{equation}
For the numerical analysis and collider simulations presented in this paper, the MPVDM model was implemented in both the {\sc\small UFO} and {\sc\small CalcHEP} formats.
The {\sc\small UFO} implementation~\cite{Degrande:2011ua, Darme:2023jdn} was generated using {\sc\small FeynRules}~\cite{Alloul:2013bka}, 
while the {\sc\small CalcHEP} implementation~\cite{Belyaev:2012qa} was produced via {\sc\small LanHEP}~\cite{Semenov:2008jy}.
Loop-level contributions relevant to the $(g-2)_\mu$ and to loop-induced kinetic mixing were computed analytically and explicitly incorporated into the CalcHEP model.
This enabled a consistent numerical evaluation of the dark matter relic density and associated cosmological constraints, which was performed using {\sc\small micrOMEGAs}~\cite{Belanger:2018ccd}.
\footnote{Both implementations are publicly available through the High Energy Physics Model Database (HEPMDB)~\cite{hepmdb}, 
with the {\sc\small UFO} model files at \href{https://hepmdb.soton.ac.uk/hepmdb:1025.0355}{https://hepmdb.soton.ac.uk/hepmdb:1025.0355} 
and the {\sc\small CalcHEP} model files and {\sc\small LanHEP} sources at \href{https://hepmdb.soton.ac.uk/hepmdb:1225.0357}{https://hepmdb.soton.ac.uk/hepmdb:1225.0357}.}


\section{$a_\mu$ in the MPVDM}
\label{sec:g-2}

The contributions to $a_\mu$ from MPVDM interactions can be either positive or negative, depending on the interplay among the new gauge, scalar, and fermion sectors.
In the \textit{tension scenario}, accommodating the observed deviation requires a net positive contribution,
whereas in the \textit{compatibility scenario} the total contribution must remain within the allowed range defined by Eq.~\eqref{eq:delta_amu_exp_c}.
The analytical expressions used to compute $a_\mu$ in the MPVDM are collected in \Cref{app:g-2analytical}.

We therefore performed a detailed numerical scan over the five-dimensional parameter space
$\{g_D, m_{V_D}, m_{\mu_D}, m_{\mu^\prime}, m_{H_D}\}$.
To ensure a uniform exploration over several orders of magnitude, the parameters were sampled uniformly in $\log_{10}$ within the following ranges:
\begin{align}
10^{-4} &\le g_D \le 4\pi, \nonumber\\
0.01~\mathrm{GeV} &\le m_{V_D} \le 10~\mathrm{TeV}, \nonumber\\
\max(100~\mathrm{GeV}, m_{V_D}) &\le m_{\mu_D} \le 10~\mathrm{TeV}, \nonumber\\
m_{\mu_D} &\le m_{\mu^\prime} \le 10~\mathrm{TeV}, \nonumber\\
0.01~\mathrm{GeV} &\le m_{H_D} \le 10~\mathrm{TeV}.
\label{eq:scan_ranges}
\end{align}

The upper bound on $g_D$ ensures perturbativity, while the lower bound avoids an excessively large relic density associated with very small couplings.
The upper bound $m_{V_D} \le 10~\mathrm{TeV}$ is motivated by the relic-density constraint discussed in the next section. As $m_{V_D}$ increases, larger values of $g_D$ are typically required to reduce the dark matter abundance to an acceptable level, and beyond this range the relic density cannot be brought into agreement with observations without violating perturbativity.
The lower bounds on the vector-like muon masses, $m_{\mu_D}$ and $m_{\mu^\prime}$, are motivated by the LEP L3 search for heavy leptons~\cite{ACHARD200175}, which sets limits around $100~\mathrm{GeV}$.
The corresponding LHC limits relevant for this study will be discussed in \Cref{sec:The LHC constraints} in the context of $\mu_D$ pair production, leading to a $\mu^+\mu^- + \slashed{E}_T$ signature.

The theoretical perturbativity and consistency requirements imposed on the scanned points are
\begin{align}
   \{\lambda, \lambda_D, y, y^\prime\} \le 4\pi,
   \qquad
   \frac{m_{V_D} - m_{V^\prime}}{m_{V_D}} < 0.5,
   \label{eq:pert_constraints}
\end{align}
where $m_{V_D}$ and $m_{V^\prime}$ are the one-loop renormalised vector masses.
These conditions exclude non-perturbative couplings and excessively large loop-induced mass splittings that would undermine the consistency of the perturbative treatment.

To quantify the agreement between the NP contribution and the experimentally preferred values, we define the dimensionless quantity
\begin{align}
   \Delta \hat{a}_\mu^i
   = \frac{\Delta a_\mu^{\mathrm{NP}} - \Delta a_\mu^{\mathrm{EXP},i}}{\sigma_i},
   \label{eq:amu_hat}
\end{align}
where the index $i=t,c$ denotes the \textit{tension} and \textit{compatibility} scenarios, respectively.
For the \textit{tension scenario}, we adopt
$\Delta a_\mu^{\mathrm{EXP,t}} = 249\times10^{-11}$
and $\sigma_t = 48\times10^{-11}$, corresponding to the 2023 experimental--theory comparison given in \Cref{eq:delta_amu_exp_t}.
For the \textit{compatibility scenario}, we follow the 2025 interpretation discussed in \Cref{sec:intro}
and require that the NP contribution remain within the combined uncertainty,
taking $\Delta a_\mu^{\mathrm{EXP,c}} = 38\times10^{-11}$
and $\sigma_c = 63\times10^{-11}$ as specified in \Cref{eq:delta_amu_exp_c}.
In this case, the criterion tests consistency with the SM--experiment agreement rather than the reproduction of a non-zero central deviation.


\subsection{Tension scenario}
\label{sec:amu_excess}

\begin{figure}[htbp]
\begin{subfigure}[b]{0.5\textwidth}
\centering
\hspace*{-18pt}\includegraphics[width=1.1\textwidth]{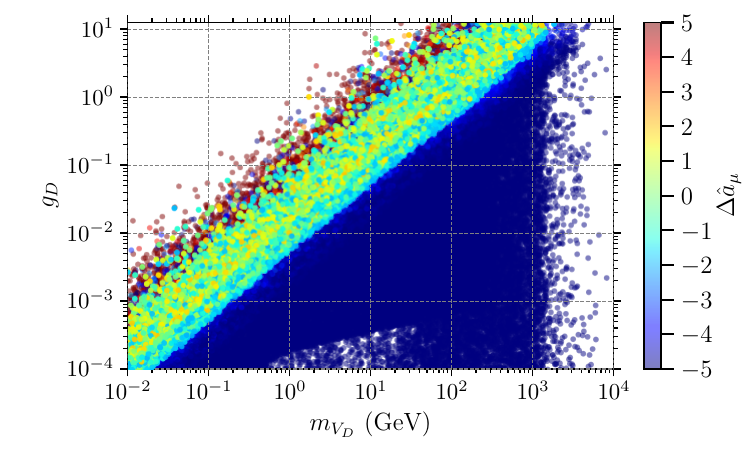}
(a)
\end{subfigure}\hfill
\begin{subfigure}[b]{0.5\textwidth}
\centering
\includegraphics[width=1.1\textwidth]{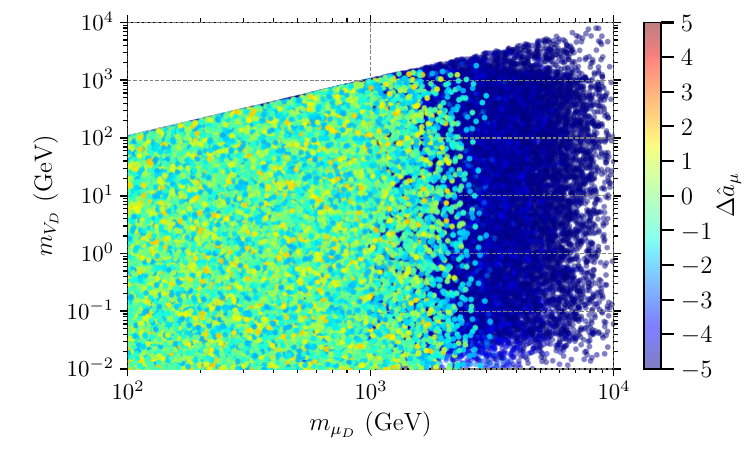}
\hspace*{20pt}(b)
\end{subfigure}
\caption{\label{fig:g-2-scatt-tension}
Colour map of $\Delta \hat{a}_\mu$ (from Eq.~\eqref{eq:amu_hat}) obtained from a five-dimensional scan of the parameter space (Eq.~\eqref{eq:pert_constraints}), projected onto the $(m_{V_D}, g_D)$ plane. The selected points reproduce the experimental value of $a_\mu$ within $5\sigma$. Perturbativity constraints from Eq.~\eqref{eq:pert_constraints} have been applied.}
\end{figure}

In this subsection we investigate the parameter space of the MPVDM model corresponding to the {"tension"} interpretation, where the experimentally measured value of $a_\mu$ exceeds the SM prediction by $\Delta a_\mu^{\mathrm{EXP}}$ defined in Eq.~\eqref{eq:delta_amu_exp_t} {by around 5$\sigma$}. 
The impact of the new physics parameters is visualised in \Cref{fig:g-2-scatt-tension}.
The colour map in~\Cref{fig:g-2-scatt-tension}(a,b) shows the deviation parameter $\Delta \hat{a}_\mu$ (defined in Eq.~\eqref{eq:amu_hat}) projected onto the $(m_{V_D}, g_D)$ and $(m_{V_D}, m_{\mu_D})$ planes. 
The points displayed satisfy the perturbativity criteria of Eq.~\eqref{eq:pert_constraints} and reproduce the experimental value of $a_\mu$ within $5\sigma$.

 The model predicts that $a_\mu$ scales approximately as $g_D^2/m_{V_D}^2$, consistent with the analytical expressions derived earlier. Points with $\Delta \hat{a}_\mu \simeq -5$ (corresponding to $a_\mu$ close to its SM value) appear in the dark-blue region of~\Cref{fig:g-2-scatt-tension}(a,b), where both $m_{V_D}$ and the masses of the vector-like muons $m_{\mu_D}, m_{\mu^\prime}$ are large (above a few TeV). In this regime, the NP contribution decouples and becomes negligible. The lighter, red-to-yellow regions indicate where the MPVDM contribution grows large enough to reproduce the experimental excess.

Of particular interest is the narrow band where $|\Delta \hat{a}_\mu| < 2$, corresponding to {\it agreement with 
an excess} at the 95\% confidence level. This region, which successfully explains the  $a_\mu$ anomaly, follows the relation $m_{V_D}/g_D \simeq 100~\mathrm{GeV}$. The five-dimensional scan thus provides an important qualitative understanding of how the relevant observables depend on the MPVDM parameters and helps to identify the viable regions that can explain the measured excess.

\begin{figure}[htbp]
\centering
\includegraphics[width=0.8\textwidth]{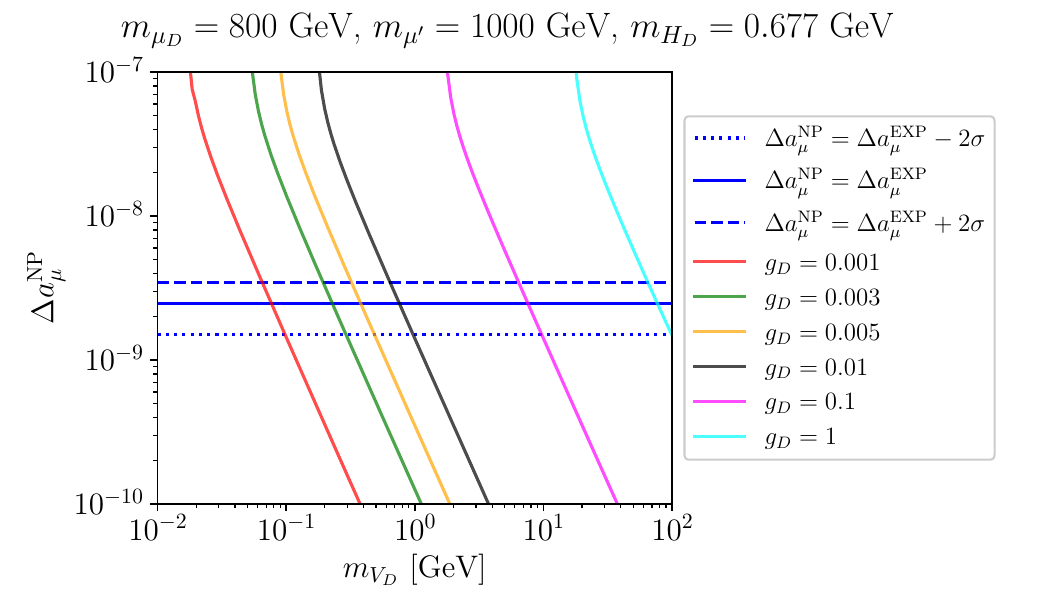}\\
\caption{\label{fig:1D-g-2-MVD}
$\Delta a_\mu^{\mathrm{NP}}$ versus $m_{V_D}$ for different values of $g_D$. 
The dotted, solid, and dashed blue lines correspond respectively to 
$\Delta a_\mu^{\mathrm{NP}} = \{\Delta a_\mu^{\mathrm{EXP}} - 2\sigma, \Delta a_\mu^{\mathrm{EXP}}, \Delta a_\mu^{\mathrm{EXP}} + 2\sigma\}$. 
Here we choose $g_D \in \{0.001, 0.003, 0.005, 0.01, 0.1, 1\}$, 
$m_{\mu_D} = 800~\mathrm{GeV}$, $m_{\mu^\prime} = 1000~\mathrm{GeV}$, and 
$m_{H_D} = 0.677~\mathrm{GeV}$, one of which corresponds to the benchmark point in~\Cref{tab:mergedBPs}.}
\end{figure}

To see the detailed numerical dependence, we examine $\Delta a_\mu^{\mathrm{NP}}$ as a function of $m_{V_D}$ for fixed values of other parameters. The result is shown in~\Cref{fig:1D-g-2-MVD}, {where the dark matter mass $m_{V_D}$ varies from 0.01 to 100~GeV,}$m_{\mu_D}=800~\mathrm{GeV}$, $m_{\mu^\prime}=1000~\mathrm{GeV}$, and $m_{H_D}=0.677~\mathrm{GeV}$, using several values of $g_D \in \{0.001, 0.003, 0.005, 0.01, 0.1, 1\}$.

In this figure, the dotted, solid, and dashed blue lines correspond to $\Delta a_\mu^{\mathrm{NP}} = \{ \Delta a_\mu^{\mathrm{EXP}} - 2\sigma$, $\Delta a_\mu^{\mathrm{EXP}}$, $\Delta a_\mu^{\mathrm{EXP}} + 2\sigma \}$, respectively. These blue lines form the 95\% CL band inside which the MPVDM model reproduces the experimental $a_\mu$ data. The parameter space given by the intersection of these lines with the blue band reproduces the experimentally measured value of $a_\mu$ in the "tension" assumption. 

For comparatively large $m_{\mu_D}$ and $m_{\mu^\prime}$ around the TeV scale (motivated by collider constraints), the dark matter mass is limited from above by perturbativity on $g_D$. For example, from~\Cref{fig:1D-g-2-MVD} one can see that $m_{V_D} \simeq 100~\mathrm{GeV}$ requires $g_D \simeq 1$, while for $m_{V_D} \simeq 1~\mathrm{GeV}$, $g_D \simeq 0.01$ suffices to match $a_\mu$ data.
The 1D dependence also reveals the role of the mass ratio $r_D = m_{\mu_D}/m_{\mu^\prime}$. As $m_{\mu^\prime}$ increases, $r_D$ decreases, and since $\Delta a_\mu^{\mathrm{NP}} \propto (1 - r_D^2)$, this enhances $a_\mu$. For instance, a small decrease in $r_D$ from 0.94 to 0.80 requires an increase of $m_{V_D}$ from about 0.14~GeV to 0.25~GeV to maintain the same $\Delta a_\mu$, illustrating the $(1 - r_D^2)$ scaling discussed analytically.

\subsection{Compatibility scenario}
\label{sec:amu_no_excess}

The updated 2025 Muon $g-2$ Theory Initiative~\cite{Aliberti:2025WP} has shown that the SM prediction for $a_\mu$ agrees with the experimental world average within the combined uncertainty. In this subsection, we explore the MPVDM parameter space corresponding to this scenario, in which $\Delta a_\mu^{\mathrm{NP}}$ must remain small enough not to spoil the SM--experiment consistency. We retain the same five-dimensional parameter scan as in the tension case but restrict to points that satisfy $|\Delta \hat{a}_\mu| < 5$.

\begin{figure}[htbp]
\begin{subfigure}[b]{0.5\textwidth}
\centering
\hspace*{-18pt}\includegraphics[width=1.1\textwidth]{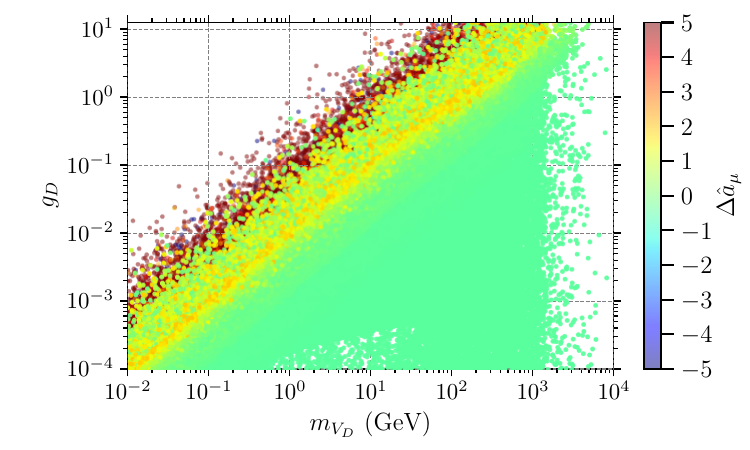}
(a)
\end{subfigure}\hfill%
\begin{subfigure}[b]{0.5\textwidth}
\centering
\includegraphics[width=1.1\textwidth]{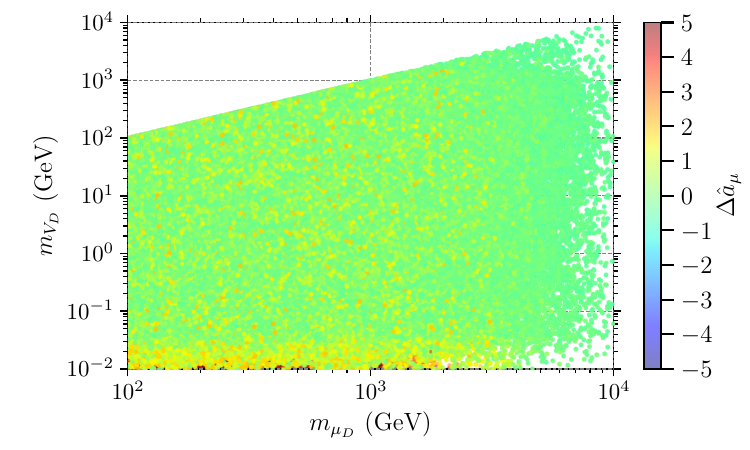}
\hspace*{20pt}(b)
\end{subfigure}
\caption{\label{fig:g-2-scatt-compatibility}
Colour map of $\Delta \hat{a}_\mu$ (from Eq.~\eqref{eq:amu_hat}) obtained from a five-dimensional scan of the parameter space (Eq.~\eqref{eq:pert_constraints}), projected onto the $(m_{V_D}, g_D)$ plane. The selected points reproduce the experimental value of $a_\mu$ within $5\sigma$ under the assumption of no $a_\mu$ excess. Perturbativity constraints from Eq.~\eqref{eq:pert_constraints} have been applied.}
\end{figure}

The qualitative structure of the parameter space remains similar to the tension case but shifts toward the decoupling regime, where the NP contributions become negligible. The dark-blue region observed in the tension scenario, corresponding to a $-5\sigma$ downward deviation of $a_\mu$ from the SM value, is absent here. While such a suppression could, in principle, occur in the presence of destructive NP interference, the MPVDM model cannot reproduce this behaviour. 

Most of the parameter space visible in~\Cref{fig:g-2-scatt-compatibility}(a,b) is consistent with the SM prediction, as expected in this scenario. It is represented by the green region corresponding to $|\Delta \hat{a}_\mu| \lesssim 2$. This regime is realised for $m_{V_D}/g_D \gtrsim 1000$~GeV and/or for large values of $m_{\mu_D}$ and $m_{\mu^\prime}$, where the NP contributions are strongly suppressed due to their $g_D^2/m_{V_D}^2$ scaling. As in the previous case, lighter $m_{V_D}$ values require smaller couplings to remain compatible with the $2\sigma$ constraint.

This demonstrates that the MPVDM model naturally accommodates the compatibility scenario without fine-tuning, as the NP effects automatically decouple when the new states are heavy and/or when $m_{V_D}/g_D$ is sufficiently large.

The boundary of the parameter space excluded at the 95\%~CL (corresponding to a $+2\sigma$ excess) follows the same approximate $m_{V_D}/g_D$ scaling as in the tension case. For $m_{V_D}/g_D > 1000$~GeV one observes $+2\sigma$, $+3\sigma$, and $+5\sigma$ deviations, which are disfavoured in the compatibility scenario.

In conclusion, both the tension and compatibility scenarios can be realised within the MPVDM framework, depending on the choice of parameters. The model remains flexible enough to reproduce the earlier $a_\mu$ anomaly or to comply with the latest experimental--theory agreement. In the next section, we will investigate how these regions intersect with cosmological and collider constraints, providing a comprehensive view of the allowed parameter space.

\section{Cosmological constraints}
\label{sec:cosmo}

In this section, we discuss the cosmological implications of the MPVDM model, focusing on the dark matter (DM) relic density, as well as the constraints from direct and indirect detection experiments. 
We have performed a scan of the model parameter space as described in the previous section and applied the latest experimental limits. 
All relic density and detection rate calculations were performed using the {\sc\small CalcHEP} version of the model (see end of \Cref{sec:model}) in the {\sc micrOMEGAs} v6.2.5 package~\cite{Belanger:2018ccd}, following the same methodology as in our previous study~\cite{Belyaev:2022shr}. 

The dark matter relic density is compared against the {\sc Planck} measurement~\cite{Planck:2018vyg},
\begin{equation}
\Omega^{\text{Planck}}_{\text{DM}}h^2 = 0.12 \pm 0.001 \;,
\label{eq:planck}
\end{equation}
and we consider as viable all parameter points that yield $\Omega_{\text{DM}}h^2 \le \Omega^{\text{Planck}}_{\text{DM}}h^2$. 
Underabundant points are interpreted as scenarios where MPVDM dark matter constitutes only a fraction of the total DM relic abundance.

In the MPVDM framework, the dark vector boson does not couple directly to light quarks or gluons at tree level. 
However, loop-induced interactions between $\gamma(Z)$ and $V_D$ are generated through kinetic mixing and triangle diagrams, as derived in~\cite{Belyaev:2022shr}, whose results we adopt here for the MPVDM realisation. 
These interactions mediate spin-independent scattering of DM off nuclei, allowing comparison with direct detection (DD) searches.

We evaluate DD constraints including the latest {\sc LUX-ZEPLIN} (LZ) 2024 limits~\cite{LZ:2024zvo}, based on 4.2 tonne-years of exposure. 
Earlier results from {\sc LZ}~\cite{LZ:2018qzl}, {\sc XENON1T}~\cite{XENON:2018voc}, and {\sc PandaX}~\cite{PandaX-4T:2021bab} are also included for completeness but are superseded by the most recent LZ constraints. 
Additionally, the {\sc DarkSide} and {\sc CRESST-III} results probing sub-GeV DM masses are included, as they constrain the light DM regime not accessible to xenon-based experiments.

In this model, DM--nucleon scattering arises entirely from loop-induced processes, dominated by triangle diagrams and $V'/Z/\gamma$ kinetic mixing. 
The corresponding low-energy interactions can be effectively described by the $V_D V_D Z$ and $V_D V_D \gamma$ vertices, with momentum-dependent form factors computed in~\cite{Belyaev:2022shr,Belyaev:2022zjx}. 
In particular, the evaluation of spin-independent DD rates from the $V_D V_D \gamma$ vertex is non-trivial, since this coupling induces a long-range force leading to divergences in standard {\sc micrOMEGAs} routines.
To handle this properly, we employ the {\tt DD\_pval} routine, which calculates the probability that the predicted DM signal is consistent with experimental data, taking into account background fluctuations. 
Parameter points with {\tt DD\_pval} $< 0.1$ are considered excluded at the 90\% confidence level (CL). 
We also use the {\tt DD\_factor} routine, which determines the multiplicative factor by which the predicted DM--nucleon cross section would need to increase to reach the 90\% CL exclusion limit. 
This exclusion factor provides a quantitative measure of proximity to current experimental sensitivity.


The indirect detection CMB constraints are evaluated using the {\tt PlanckCMB(sigmaV, SpA, SpE)} subroutine. 
Constraints derived from {\sc Planck} data rely on the fact that DM annihilation into Standard Model (SM) particles can inject significant energy into the photon--baryon plasma during recombination. 
The resulting energy deposition modifies the ionisation history and the CMB anisotropy spectrum, leading to an upper bound on the DM annihilation power~\cite{Planck:2018vyg}:
\begin{equation}
   P_{\text{ann}} <  P^\text{PLANCK}_{\text{ann}} = 3.2\times10^{-28}\,\frac{\mathrm{cm^3}}{\mathrm{s\,GeV}} \quad \text{at 95\% C.L.,}
\text{\ \ \ where\ \ }
   P_{\text{ann}} = \sum_j \frac{f^{\mathrm{eff}}_j \langle\sigma v\rangle_j}{m_{\mathrm{DM}}}
   \left( \frac{\Omega_{\mathrm{DM}}}{\Omega_{\mathrm{DM}}^{\mathrm{Planck}}} \right)^2 .
   \label{eq:Pann_def}
\end{equation}
Here, $\langle\sigma v\rangle_j$ is the thermally averaged annihilation cross section for channel $j$, and $f^{\mathrm{eff}}_j$ denotes the fraction of the annihilation energy absorbed by the plasma for that channel, studied in~\cite{Slatyer:2015jla,Leane:2018kjk}. 
The factor $(\Omega_{\mathrm{DM}}/\Omega_{\mathrm{DM}}^{\mathrm{Planck}})^2$ accounts for the reduced annihilation rate when the relic abundance of MPVDM dark matter is below the observed value.

\begin{figure}[htbp]
\centering
\includegraphics[width=0.8\textwidth]{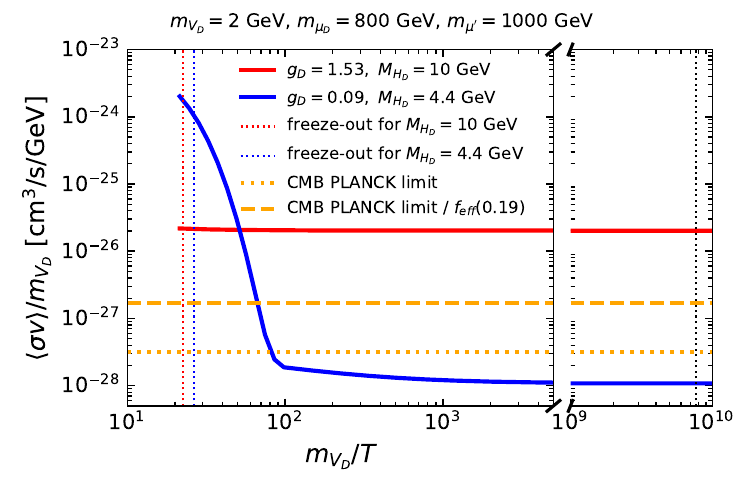}
\caption{\label{fig:CMB} 
Evolution of the thermally averaged annihilation rate $\langle\sigma v\rangle/m_{\mathrm{DM}}$ as a function of $m_{\mathrm{DM}}/T$ from the freeze-out epoch to the CMB epoch for representative MPVDM benchmark points. 
See text for a detailed description of the curves.}
\end{figure}

In the MPVDM model, the low-mass annihilation $V_D V_D \to V^\prime V^\prime$ receives contributions from multiple topologies, including $s$-channel scalar exchange via $H$ and $H_D$ as well as $t/u$-channel vector exchange. 
A potential near-resonant enhancement is also possible, controlled by the condition $2m_{\mathrm{DM}}\simeq m_{H_D}$, which produces an important \textit{kinematical} effect that suppresses the annihilation rate at the CMB epoch, as discussed below.

For DM masses around or below 10~GeV, a pure $s$-wave annihilation mechanism is excluded: the annihilation rate that determines the relic abundance at freeze-out remains essentially unchanged at the CMB epoch, so $\langle\sigma v\rangle/m_{\mathrm{DM}}$ stays large and violates the {\sc Planck} bound. 
However, in the resonant annihilation case the situation is drastically different. 
If $2m_{\mathrm{DM}}$ lies slightly below $m_{H_D}$, the thermal motion of DM at freeze-out shifts the center-of-mass energy onto the resonance, enhancing $\langle\sigma v\rangle$. 
At the CMB epoch, when DM velocities are much lower, the system moves off resonance and $\langle\sigma v\rangle$ is suppressed by several orders of magnitude. 
This near-resonant regime typically requires $2m_{\mathrm{DM}}$ to be below about 10--20\% of $m_{H_D}$ and does not involve any fine-tuning of parameters. 
Consequently, $\langle\sigma v\rangle/m_{\mathrm{DM}}$ can drop dramatically between freeze-out ($m_{\mathrm{DM}}/T \sim 10$) and the CMB epoch ($m_{\mathrm{DM}}/T \sim 10^{10}$), as illustrated in~\Cref{fig:CMB}.

The red solid line in~\Cref{fig:CMB} corresponds to an off-resonant configuration with $(m_{\mathrm{DM}}, m_{H_D}) = (2, 10)$~GeV, where $\langle\sigma v\rangle/m_{\mathrm{DM}}$ remains nearly constant over time and exceeds the CMB limit (orange dashed line). 
In contrast, the blue solid line shows the near-resonant case $(m_{\mathrm{DM}}, m_{H_D}) = (2, 4.4)$~GeV, where $\langle\sigma v\rangle/m_{\mathrm{DM}}$ decreases by about four orders of magnitude between freeze-out and the CMB epoch, falling well below the {\sc Planck} upper bound once divided by the energy-injection efficiency factor $f_{\mathrm{eff}} = 0.19$. 

This resonance-driven {\it kinematical suppression} mechanism provides a generic way to evade the CMB constraint, allowing even sub-GeV dark matter to remain consistent with cosmological observations while reproducing the correct relic abundance. 
In the MPVDM and similar near-resonant scenarios, the suppression arises simply because the dark matter temperature follows that of the Standard Model plasma: as the Universe cools, the dark matter velocity distribution shifts to lower values, moving the annihilation process progressively off the resonance. 
This happens without any need for early kinetic decoupling or additional model-dependent effects. 
In contrast, the ``belated freeze-out'' picture of Ref.~\cite{Belanger:2025kce} relies on dark matter decoupling kinetically from the plasma and cooling faster than the CMB, with ultra-narrow Breit--Wigner resonances required to maintain a sufficient suppression at late times. 
While Ref.~\cite{Belanger:2025kce} also noted that off-resonance annihilation at recombination can reduce $\langle\sigma v\rangle$, we find that this {\it purely thermal and kinematical detuning}---without invoking ultra-narrow widths or early kinetic decoupling---is already sufficient to satisfy the {\sc Planck} bound once realistic final states and energy-injection efficiencies are included. 
The MPVDM realisation explicitly illustrates this behaviour, highlighting a generic, previously overlooked near-resonant regime in which standard thermal dark matter can evade CMB constraints purely through its temperature-driven kinematical evolution, an essential feature of the MPVDM framework in the light dark matter regime.


\begin{figure}[htbp]
\centering\includegraphics[trim={0 6.1cm 0 0},clip,width=0.8\textwidth]{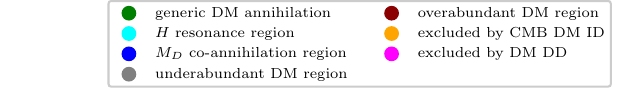}
\begin{subfigure}[b]{0.495\textwidth}
\includegraphics[trim={0 0 0 1.5cm},clip,width=\textwidth]{fig/cosmo_MVP_GD.pdf}
        \centering\small{(a)}
\end{subfigure}
\begin{subfigure}[b]{0.495\textwidth}
\includegraphics[trim={0 0 0 1.5cm},clip,width=\textwidth]{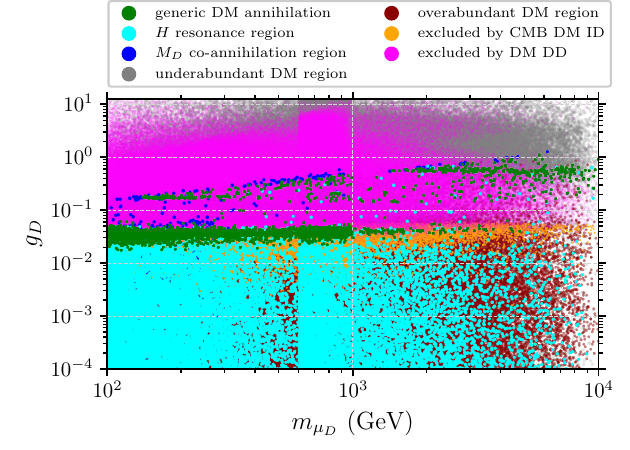}
        \centering\small{(b)}
\end{subfigure}\\
\begin{subfigure}[b]{0.495\textwidth}
\includegraphics[trim={0 0 0 1.5cm},clip,width=\textwidth]{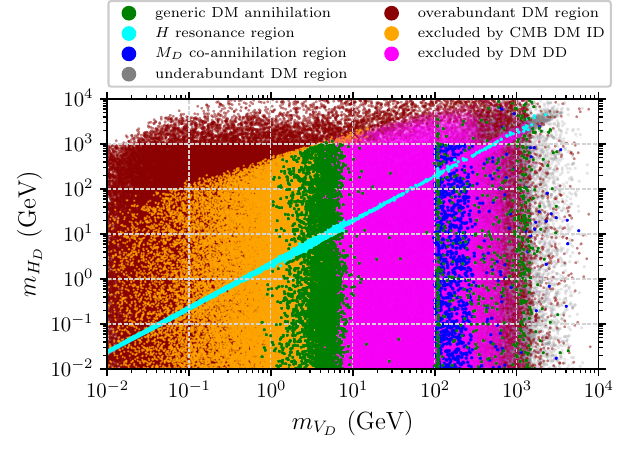}
        \centering\small{(c)}
\end{subfigure}
\begin{subfigure}[b]{0.495\textwidth}
\includegraphics[trim={0 0 0 1.5cm},clip,width=\textwidth]{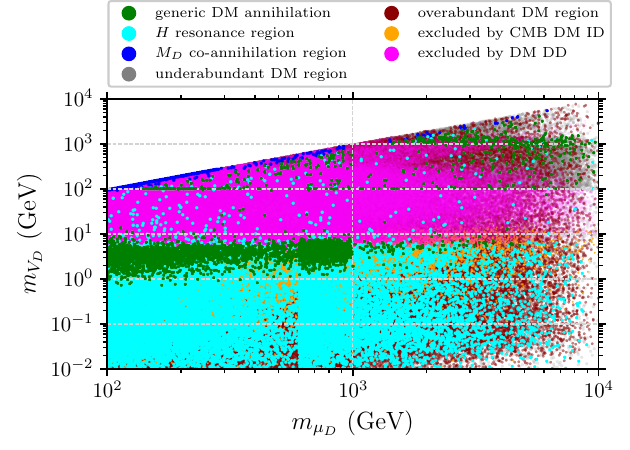}
        \centering\small{(d)}
\end{subfigure}
\caption{\label{fig: per constraints + cosmo} 
Results from the 5D scan in different projections: 
(a) $(m_{V_D}, g_D)$, 
(b) $(m_{\mu_D},g_D)$, 
(c) $(m_{V_D},m_{H_D})$, and 
(d) $(m_{\mu_D},m_{V_D})$ planes, respectively. 
The perturbativity and cosmological constraints have been applied on each individual panel. 
The cosmological limits include (1) DM relic density, (2) DM DD, and (3) DM ID. 
The allowed regions are coloured green, cyan, blue, and grey, while the excluded ones are highlighted in dark red, orange, and magenta. 
The white region corresponds to violation of the perturbativity constraint.}
\end{figure}

In~\Cref{fig: per constraints + cosmo}, we show the regions allowed or excluded by various observables in 2D planes corresponding to different projections of the model parameter space. 
The green, cyan, and blue regions are allowed by perturbativity, DD, and ID constraints and yield a relic density $\Omega_{\rm DM}^{\text{Planck}}h^2 \pm 0.012$. 
The grey region also satisfies these constraints but corresponds to an under-abundant relic density. 
In contrast, the dark red, orange, and magenta regions are excluded by the relic density, CMB ID, and DD limits, respectively. 
The green region, labelled as \textit{generic DM annihilation}, appears as a small diagonal strip in~\Cref{fig: per constraints + cosmo}(a), where the dominant annihilation channels are $V_D V_D^* \to V^\prime V^\prime$. 
However, these generic DM annihilation processes are not efficient enough and are excluded by the DM ID constraint in the region of small DM masses below 1~GeV and small couplings $g_D<0.02$. 
One can see that the green strip is not uniform over the range $1<m_{V_D}/\text{GeV}<10^4$, especially in the region $10<m_{V_D}/\text{GeV}<100$, where the DM DD constraint becomes relatively stronger. 
Regions above (below) the green strip correspond to over- (under-) abundant relic densities, indicated in dark red and grey, respectively. 
The cyan region corresponds to the $H_D$ \textit{resonance region}, where the main annihilation channel is $V_D V_D^* \to H_D \to V^\prime V^\prime$. 
This occurs when the DM mass approaches half of $m_{H_D}$. 
The resonance appears as a diagonal cyan strip in~\Cref{fig: per constraints + cosmo}(c). 

Finally, the blue region corresponds to the $\mu_D$ \textit{co-annihilation region}, which occurs when the DM mass approaches $m_{\mu_D}$, as seen in Figs.~\ref{fig: per constraints + cosmo}(b) and (d). 
The co-annihilation process becomes significant when the DM mass reaches $\sim100$~GeV (the lower limit for vector-like muons from LEP). 
For small couplings $g_D < 0.1$, co-annihilation proceeds mainly via $\mu_D\mu_D \to q\bar{q}, \ell^+\ell^-, \nu\bar{\nu}$ through photon and $Z$ exchange, which is independent of $g_D$ and occurs in the region $100<m_{V_D}/\text{GeV}<300$. 
When the coupling $g_D$ increases, the dominant channel shifts to $V_D \mu_D \to \gamma\mu, \gamma \mu^\prime$ via $\mu$ or $\mu^\prime$ exchange.

\section{Collider Constraints}
\label{sec:collider pheno}

In this section, we discuss collider limits on the MPVDM model based on LHC data by reinterpreting ATLAS and CMS searches in the context of vector-like (VL) muons predicted by the MPVDM model and also highlight the novel multi-lepton signatures with up to six or more muons in the final state. 
The detailed properties of representative benchmark points  that satisfy all relevant constraints, including $a_\mu$, cosmology, and collider limits, are collected in the combined benchmark Table~\ref{tab:mergedBPs} in Section~\ref{sec:benchmarks}, where we provide branching ratios, production cross sections, and expected event yields for the most striking signatures.

\begin{figure}[htbp]
\centering
\begin{minipage}{.41\textwidth}
\includegraphics[width=\textwidth]{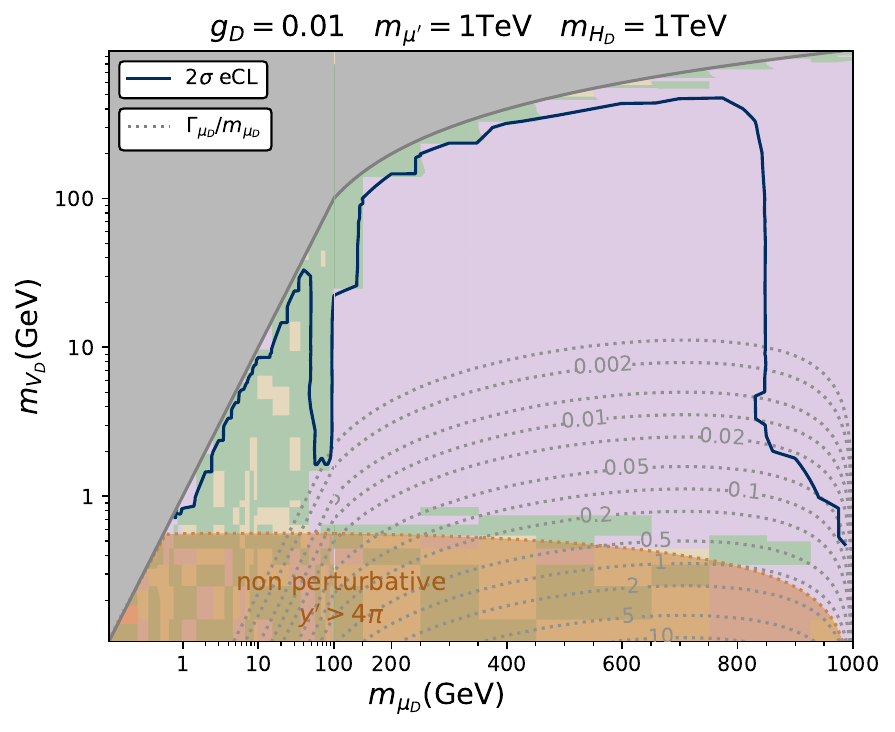} \\
\includegraphics[width=\textwidth]{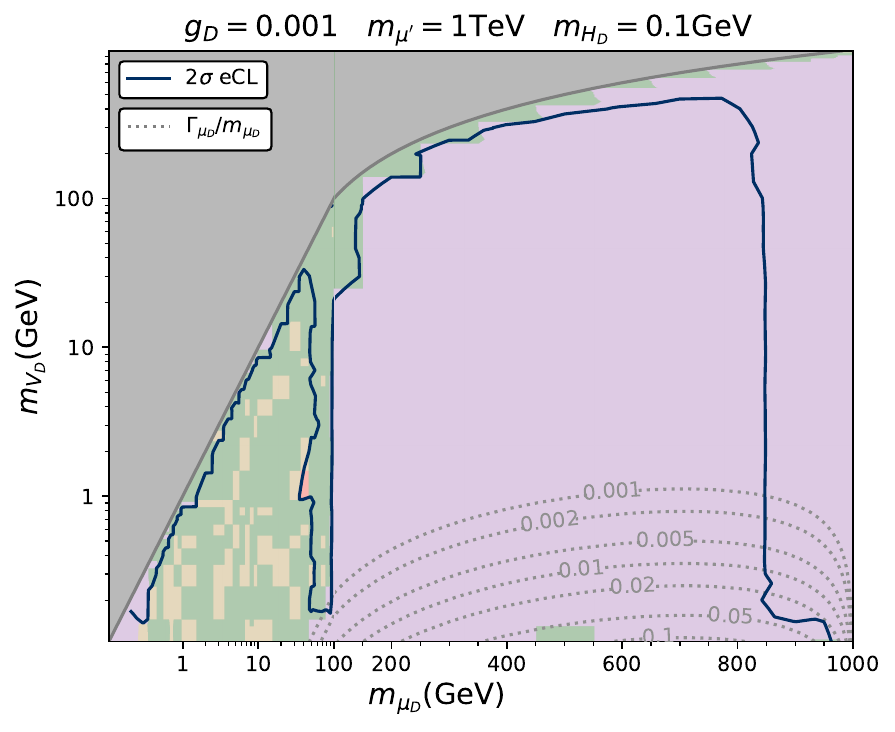}
\end{minipage}%
\begin{minipage}{.41\textwidth}
\includegraphics[width=\textwidth]{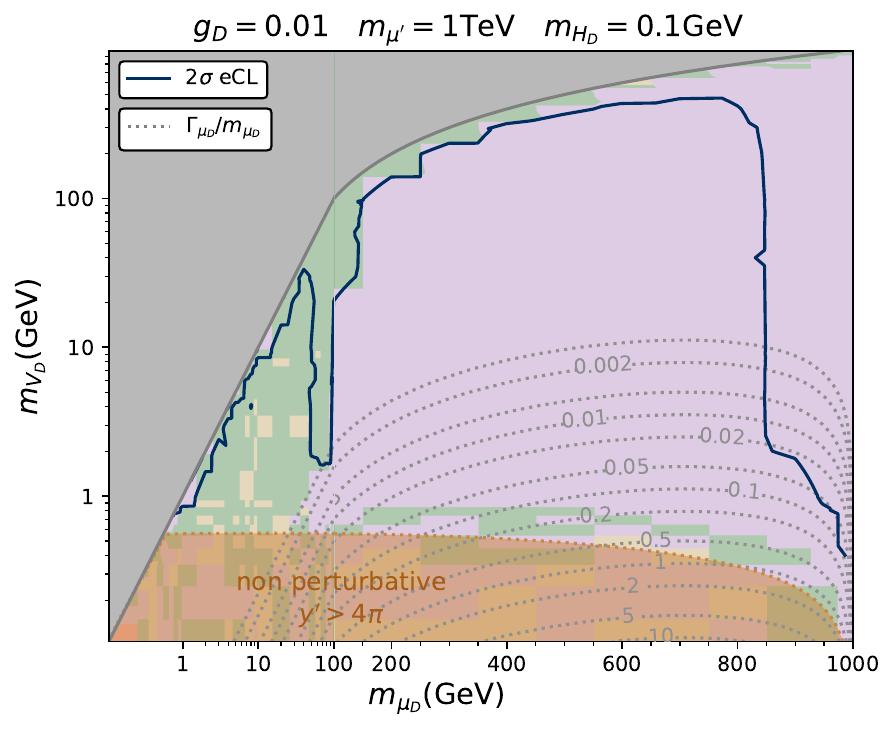}\\
\includegraphics[width=\textwidth]{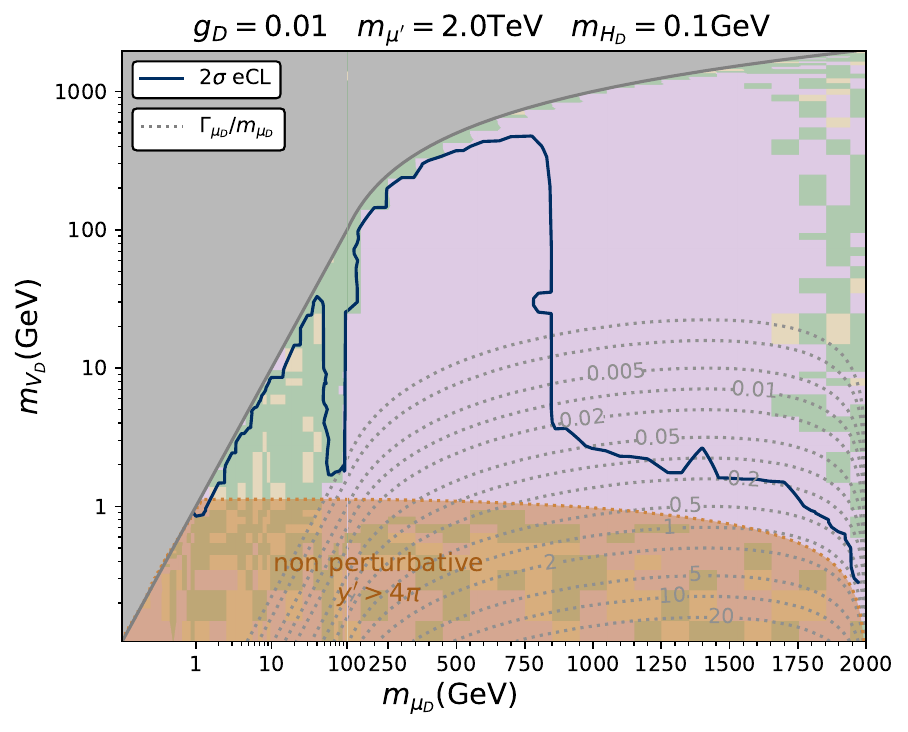}
\end{minipage}%
\begin{minipage}{0.18\textwidth}
\includegraphics[width=\textwidth]{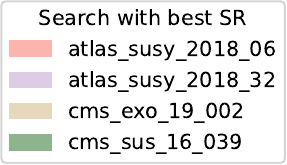}
\end{minipage}
\caption{\label{fig:recasting}
Mass limits in the $(m_{\mu_D},m_{V_D})$ plane for $pp \to \mu_D^+ \mu_D^- \to \mu^+ \mu^- + \slashed{E}_T$ based on a recast of ATLAS and CMS searches. 
Simulations have been performed for $g_D=0.001,0.01$, $m_{\mu^\prime}=1000,2000$~GeV, and $m_{H_D}=0.1,1000$~GeV. 
The dark blue line indicates the exclusion limit at 95 percent C.L., and the dotted lines show the ratio of the $\mu_D$ decay width to its mass. 
The background colour indicates the search providing the most sensitive signal region that determines the exclusion limit. 
The area where non-perturbative couplings are obtained is shaded in orange.}
\end{figure}

\subsection{Lower mass limits on vector-like muons from \texorpdfstring{$pp \to \mu^+ \mu^- + \slashed{E}_T$}{Lg}}
\label{sec:The LHC constraints}

Vector-like muons $\mu_D^+\mu_D^-$ are produced at the LHC via $\gamma$, $Z$, and $V^\prime$ exchange. 
The first two channels dominate because $\mu_D$ carries the same hypercharge as the SM muon, while production via $V^\prime$ is highly suppressed by the small kinetic mixing. 
After production, each $\mu_D$ decays entirely into a SM muon and a dark vector boson $V_D$ through Yukawa-induced mixing between SM and VL muons.

Other processes, such as pair production of $\mu^\prime$ or pair and associate production of $H_D$ and $V^\prime$ do not give competitive bounds with respect to the aforementioned process using the searches currently available at the LHC: the former because of the hierarchy between masses in the fermion sector, which force $\mu^\prime$ to be the heaviest of the three fermions, thus making the cross-section for pair production always smaller than the one of $\mu_D$ (but see next section for the possibility of striking signatures from this process which can be probed at the HL-LHC), the latter again because of the smallness of the cross-section (see also our previous studies~\cite{Belyaev:2022shr,Belyaev:2022zjx}).

To obtain LHC limits on our model we reinterpret an ATLAS search for slepton pair production, ATLAS-SUSY-2018-32~\cite{ATLAS:2019lff}, $pp \to \tilde{\ell} \tilde{\ell}\to \mu^+ \mu^- + \slashed{E}_T$, as constraints on the analogous MPVDM process $pp \to \mu_D^+ \mu_D^- \to \mu^+\mu^- + \slashed{E}_T$. 
Furthermore, we recast searches featuring multi-leptons and $\slashed{E}_T$ in the final state, namely CMS-SUS-16-039~\cite{CMS:2017moi}, designed to target electroweakinos, and CMS-EXO-19-002~\cite{CMS:2019lwf}, targeting type III see-saw and top-philic scalars, which are sensitive to our signal, especially in the low mass region. 
By recasting these analyses, we derive exclusion regions at 95 percent C.L., which allow us to set a lower bound on the VL muon masses.

The signal has been simulated using \mg~\cite{Alwall:2014hca} using the {\sc\small UFO} version of the model (see the end of \Cref{sec:model}) and the LO NNPDF4.0 parton distribution functions~\cite{NNPDF:2021njg, Buckley:2014ana}. To account for finite widths and interference effects, the simulations have been performed without imposing that $\mu_D$ is resonant, but going directly to the $\mu^+\mu^-+2 V_D$ final state. The showering and hadronisation of the parton-level events have been performed through \pythia~\cite{Sjostrand:2014zea}. The recast of experimental searches has been done through the {\sc\small MadAnalysis 5} framework~\cite{Conte:2012fm, Conte:2014zja, Conte:2018vmg}, which internally takes care of the detector effects through {\sc\small Delphes~3}~\cite{deFavereau:2013fsa}. The recasts of the experimental searches used in this analysis are available in the {\sc\small MadAnalysis 5} Public Analysis Database. \\ 
The results of the recast are shown in \Cref{fig:recasting} in the $(m_{\mu_D}, m_{V_D})$ plane for representative parameter choices $g_D = 0.001, 0.01$, $m_{\mu^\prime}=1000,2000$~GeV, and $m_{H_D}=0.1,1000$~GeV which broadly cover the coupling and mass spectrum. In the phenomenologically interesting region with $m_{V_D}<1$~GeV, the lower bound on the vector-like muon mass is at least approximately 850~GeV, but large-width effects can significantly raise the limit for lower $V_D$ masses. These effects, which start to be significant when $\Gamma_{V_D}/m_{V_D}$ exceeds about 2 percent, depend on the $g_D$ coupling: this is clearly visible by comparing the top-right and bottom-left panels of \Cref{fig:recasting}, which only differ by the value of $g_D$. 
The mass of $H_D$ does not affect these bounds, since in our full $2\!\to\!4$ simulations of $pp\!\to\!\mu^+\mu^-V_DV_D$, the scalar $H_D$ contributes only through suppressed topologies that do not originate from the genuine $2\!\to\!2$ production process. 
These configurations correspond effectively to $2\!\to\!3$-type processes, where $H_D$ is radiated from internal $\mu_D$, $V'$ or $V_D$ lines, and therefore have a negligible impact on the total rate.

This bound is always driven by the slepton search~\cite{ATLAS:2019lff}, which loses sensitivity around $m_{\mu_D}=100$ GeV (corresponding to the LEP limit~\cite{ACHARD200175}). The low mass region (indeed excluded by LEP) is however also excluded by the CMS multi-lepton searches~\cite{CMS:2017moi} and~\cite{CMS:2019lwf}. As it is always the case in models which feature a $t$-channel mediator decaying to a SM particle and missing transverse energy, the small mass-gap region between $\mu_D$ and $V_D$ is not excluded due to the softness of the SM objects in the final state, which sizably decreases the sensitivity of searches looking for $\slashed{E}_T$ excesses. Here, $\slashed{E}_T$ denotes the missing transverse energy arising from undetected dark matter  particles or neutrinos in the final state. We finally notice that the LHC bounds always exclude regions where the Yukawa couplings become non-perturbative, making the recast results robust against higher-order corrections in the new couplings of the theory.  
Consistently with these limits, all four benchmarks listed in Table~\ref{tab:mergedBPs} have $m_{\mu_D}\gtrsim900$ GeV, 
together with values of $g_D$ of order $10^{-3}$ and dark-matter masses above about $0.2$ GeV, ensuring that they satisfy the relic-density and other cosmological constraints while evading the direct LHC bounds.  
At the same time, these benchmark points yield non-negligible $\mu_D^+\mu_D^-$ and $\mu^{\prime+}\mu^{\prime-}$ pair-production rates at the HL-LHC.

\subsection{Multi-lepton signatures}
\label{sec:multileptons}

The MPVDM model predicts striking multi-lepton final states, which provide distinctive signatures for collider searches. 
At the LHC, multi-lepton events can arise from the pair production of heavy vector-like muons $\mu^{\prime+}\mu^{\prime-}$. 
Each $\mu^\prime$ can decay through several channels:
1. $\mu^\prime \to \mu_D V_D$,\ \ \ \ \
2. $\mu^\prime \to \mu H_D$,\ \ \ \ \ and\ \ \ \ \
3. $\mu^\prime \to \mu V^\prime$.
\\
The scalar $H_D$ further decays into $V_D V_D^*$, $V^\prime V^\prime$, or $\mu^+\mu^-$, while $V^\prime$ subsequently decays into a muon pair. 
These cascades lead to final states with at least six muons from $pp \to \mu^{\prime+}\mu^{\prime-}$ pair production when the $V' \to \mu^+\mu^-$ decay is  kinematically open.

\Cref{tab:mergedBPs} in Section~\ref{sec:benchmarks} summarises four representative benchmark points that illustrate these signatures quantitatively. BP1 and BP2 (left block of the table) correspond to short-lived scenarios, in which the dark photon $V^\prime$ decays promptly, and differ by whether the $(g-2)_\mu$ constraint is interpreted in the tension (BP1) or compatibility (BP2) scenario; BP3 and BP4 (right block) describe long-lived $V'$ scenarios with $V'$ mass  below the dimuon threshold, $m_{V^\prime}<2m_\mu$. 
For each benchmark we list the branching ratios of $\mu^\prime$, $H_D$ and $V^\prime$, as well as the inclusive probabilities $P(\mu^\prime \mu^\prime \to \text{leptons})$ and the corresponding event yields at $\sqrt{s}=14$ TeV and an integrated luminosity of 3000 fb$^{-1}$.

Inspecting Table~\ref{tab:mergedBPs}, the probabilities for obtaining six, eight, and ten muons from $\mu^{\prime+}\mu^{\prime-}$ production lie in the ranges
\begin{align}
P(\mu^\prime \mu^\prime \to 6\mu) & \simeq 0.29\text{--}0.49,\\
P(\mu^\prime \mu^\prime \to 8\mu) & \simeq 0.11\text{--}0.15,\\
P(\mu^\prime \mu^\prime \to 10\mu) & \simeq 0.01\text{--}0.02,
\end{align}
across BP1--BP2. The corresponding event yields for the HL-LHC, also given in Table~\ref{tab:mergedBPs}, are $N_{\mathrm{event}}(pp \to 6\mu) \sim 44$, $N_{\mathrm{event}}(pp \to 8\mu) \sim 14\text{--}17$, and $N_{\mathrm{event}}(pp \to 10\mu) \sim 1\text{--}3$, depending on the benchmark. 
Since the scalar $H_D$ is highly boosted, its decay products 
($H_D\!\to\!V^\prime V^\prime$) inherit this boost, and the subsequently produced $V^\prime$ bosons are boosted as well. Their decays into $\mu^+\mu^-$ therefore yield very collimated muon pairs that can appear as merged muon jets, requiring dedicated reconstruction strategies.

In both BP3 and BP4, the mass of the dark vector $V'$ lies below the dimuon 
threshold,\footnote{A similar situation may arise even when $m_{V_D}$ is above, 
but close to, the dimuon threshold: for sufficiently heavy $\mu'$, e.g. 
$m_{\mu'}>1.5$~TeV with $m_{\mu_D}=750$~GeV, radiative corrections can shift 
$m_{V'}$ below $2m_\mu$.}
so that $V'$ behaves as a special dark photon whose decays proceed through its 
small effective kinetic mixing with the visible sector. In this regime, the only 
open decay channels are into displaced $e^+e^-$ pairs (with a branching ratio of 
order 30\%) or into neutrinos (about 70\%), the latter leading to missing 
transverse energy. The resulting displaced electron pairs — or invisible 
signatures when $V'\to\nu\nu$ — provide distinctive long-lived LHC signatures 
that complement the prompt muons from the primary $\mu'$ decays.

In general, the MPVDM model predicts several \textit{novel} and experimentally distinctive collider signatures that set it apart from other dark-sector scenarios:
\begin{itemize}
    \item[1)] Two isolated prompt muons accompanied by two pairs of boosted muons from two $V'$ decays (six muons in total);
    \item[2)] Two isolated prompt muons accompanied by one pair of boosted muons from $V'$ decay  and one cluster of four boosted muons from $H_D\to V'V'$ (eight muons in total);
    \item[3)] Two isolated prompt muons accompanied by two clusters of four boosted muons (ten muons in total);
    \item[4)] Two isolated prompt muons accompanied by two  merged displaced electrons from long-lived $V'$ decays and missing transverse momentum form   $V'$ decay to neutrinos  (two muons plus two dispalced  electrons plus MET);
    \item[5)] Two isolated prompt muons accompanied by two clusters of two displaced boosted electrons (two muons plus two pairs of displaced  electrons). This signature can also be unaccompanied missing transverse momentum from neutrinos originating from   $V'$ decay.
\end{itemize}

These novel signatures provide striking experimental targets for the LHC and future colliders, uniquely characterising the MPVDM framework and offering clear avenues for discovery and will be explored in a separate dedicated study.
The benchmarks in Table~\ref{tab:mergedBPs} furnish concrete targets for such searches, spanning both prompt and long-lived dark-photon regimes and covering the tension and compatibility interpretations of $(g-2)_\mu$.

\section{The combined sensitivity to MPVDM parameter space and benchmarks}
\label{sec:combined}

In the previous sections we have discussed the $(g-2)_\mu$, cosmological and collider constraints separately. 
Here we present a combined analysis that identifies the regions of the MPVDM parameter space simultaneously consistent with all constraints: perturbativity, collider searches, cosmology, and the muon anomalous magnetic moment. 
As usual, for the $(g-2)_\mu$ case we consider both the {\it tension scenario}, in which a positive $\Delta a_\mu$ is required, and the {\it compatibility scenario}, in which the experimental measurement agrees with the SM prediction within uncertainties. 
The full parameter scan is performed in the five-dimensional space 
$\{g_D, m_{V_D}, m_{\mu_D}, m_{\mu^\prime}, m_{H_D}\}$ with perturbativity and mass hierarchy conditions from~\Cref{eq:pert_constraints}.
When interpreting the viable regions, we will occasionally refer to the four representative benchmark points collected in the combined Table~\ref{tab:mergedBPs} of Section~\ref{sec:benchmarks} , which illustrate typical short-lived and long-lived scenarios in both the tension and compatibility frameworks.

\subsection{Tension scenario}
\label{sec:combined_tension}

\begin{figure}[htb]
\centering
\includegraphics[trim={0 6.15cm 0 0},clip,width=0.8\textwidth]{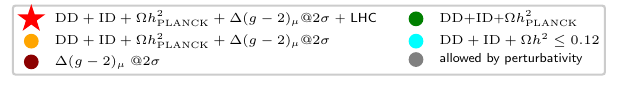}\\
\begin{subfigure}[b]{0.495\textwidth}
   \includegraphics[trim={0 0 0 1.5cm},clip,width=\textwidth]{fig/cosmo+g2+lhc_MVP_GD.pdf}\subcaption{}
\end{subfigure}
\begin{subfigure}[b]{0.495\textwidth}
   \includegraphics[trim={0 0 0 1.5cm},clip,width=\textwidth]{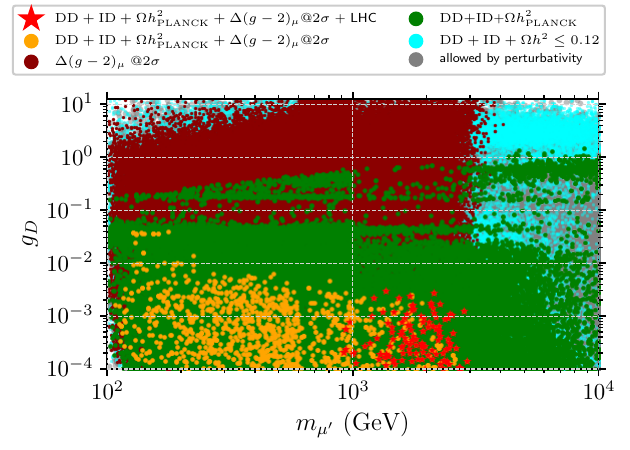}\subcaption{}
\end{subfigure}\\
\begin{subfigure}[b]{0.495\textwidth}
   \includegraphics[trim={0 0 0 1.5cm},clip,width=\textwidth]{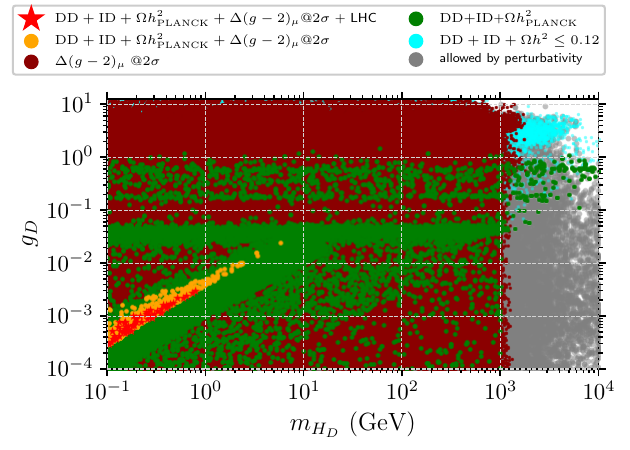}\subcaption{}
\end{subfigure}
\begin{subfigure}[b]{0.495\textwidth}
   \includegraphics[trim={0 0 0 1.5cm},clip,width=\textwidth]{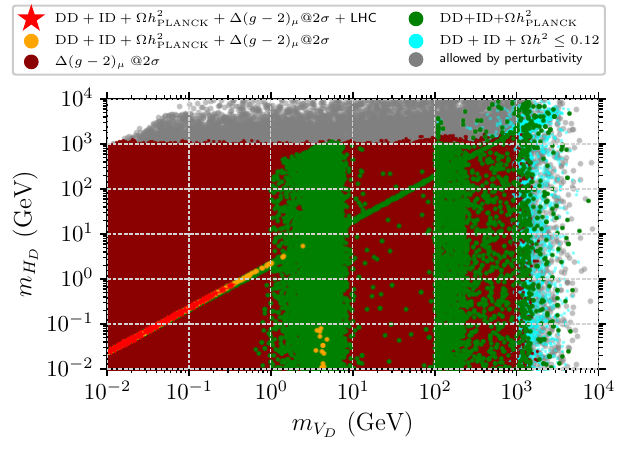}\subcaption{}
\end{subfigure}\\
\begin{subfigure}[b]{0.495\textwidth}
   \includegraphics[trim={0 0 0 1.5cm},clip,width=\textwidth]{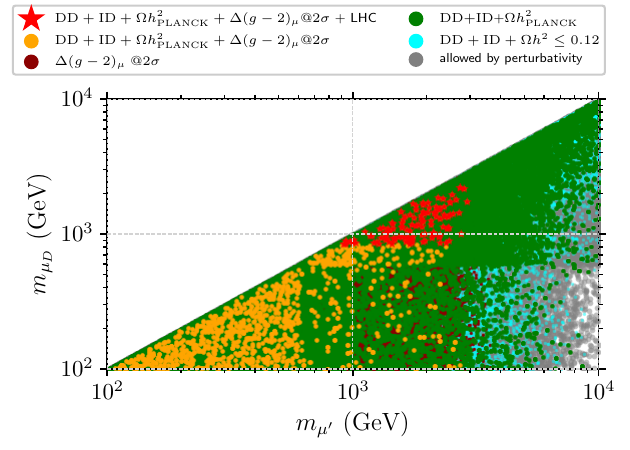}\subcaption{}
\end{subfigure}
\begin{subfigure}[b]{0.495\textwidth}
   \includegraphics[trim={0 0 0 1.5cm},clip,width=\textwidth]{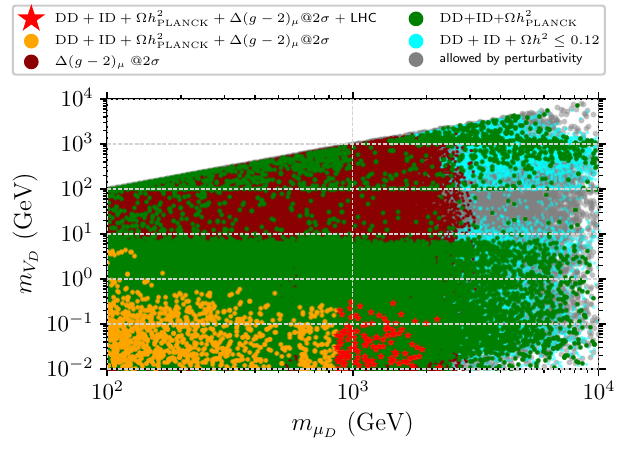}\subcaption{}
\end{subfigure}
\caption{\label{fig:scatter-cosmo-g2-lhc}
Scatter plots from the 5D parameter scan projected onto 
(a) $(m_{V_D}, g_D)$, (b) $(m_{\mu_D}, g_D)$, 
(c) $(m_{H_D}, g_D)$, (d) $(m_{H_D}, m_{V_D})$, 
(e) $(m_{\mu^\prime}, m_{\mu_D})$, and (f) $(m_{\mu_D}, m_{V_D})$ planes. 
The red points indicate regions consistent with all constraints, including perturbativity, collider limits, cosmological bounds, and $a_\mu$ (within the {\it tension scenario} at the $2\sigma$ level). 
Grey points satisfy only perturbativity. 
Green and cyan points correspond to cosmologically allowed regions (matching or below the Planck relic densityDark-red points satisfy the $a_\mu$ constraint alone, while orange points represent the parameter space consistent with the combined cosmological and $a_\mu$ constraints.}
\end{figure}

In \Cref{fig:scatter-cosmo-g2-lhc} we show the projections of the 5D scan for the {\it tension scenario}, where the positive $\Delta a_\mu$ excess is assumed.  
Each panel corresponds to a different pair of model parameters, and the colour code represents the successive application of theoretical and experimental constraints.

The grey points indicate regions that satisfy only perturbativity, while the green and cyan regions correspond to points allowed by cosmological limits, depending on whether the relic density matches the Planck measurement or is underabundant.  
Dark-red points fulfil the $a_\mu$ constraint within $2\sigma$, and orange points indicate the overlap of cosmological and $a_\mu$ constraints.  
The red stars highlight the region simultaneously consistent with all constraints: perturbativity, cosmology, $a_\mu$, and collider bounds.
Benchmark points BP1 and BP3 in Table~\ref{tab:mergedBPs} provide explicit parameter choices within this combined red region for the short-lived and long-lived cases respectively.

In the $(m_{V_D}, g_D)$ and $(m_{H_D}, g_D)$ planes, one can clearly see that the viable red region lies along the $H_D$ resonance line, corresponding to $2m_{\mathrm{DM}}\simeq m_{H_D}$.  
In this band, proximity to the $H_D$ resonance ensures that the thermal relic density is sufficiently reduced at freeze-out while remaining consistent with CMB constraints for such low dark matter masses. 
At the same time, the light $m_{\mathrm{DM}}$ values required by this mechanism naturally enhance $\Delta a_\mu$ in the MPVDM loops. 
Since the other particles entering the loop, $m_{\mu^\prime}$ and $m_{\mu_D}$, are pushed to the TeV scale by collider bounds, the light dark matter mass compensates for the suppression from heavy states and provides a large enough $\Delta a_\mu$ to explain the observed tension.

It is worth emphasising that the viable points in the 5D scan do not form a broad, structureless cloud but rather a thin, correlated hypersurface. 
This hypersurface emerges because the conditions for the correct relic density, CMB safety, and the size of the $(g-2)_\mu$ loop contribution impose mutually non-trivial relations among $m_{V_D}$, $g_D$, and the heavy-lepton masses. 
For instance, the requirement that the annihilation cross section be enhanced only during freeze-out but suppressed at late times selects points satisfying both $m_{H_D} \simeq 2m_{V_D}$ and a sufficiently small $g_D$ so that Sommerfeld and off-resonance effects do not spoil the CMB constraint. 
At the same time, the $(g-2)_\mu$ loop prefers light $m_{V_D}$ and sizable yet perturbative $g_D$, which is why the overlapping region is geometrically narrow. 
Benchmark BP1 captures this fine balance: its position along the near-resonant strip is not accidental, but a direct consequence of these multi-constraint correlations.

The preferred range of parameters is 
\[
10^{-4} \lesssim g_D \lesssim 3\times10^{-3}, \quad 0.01~{\rm GeV} \lesssim m_{V_D} \lesssim 1~{\rm GeV}, \quad m_{H_D}\lesssim 1~{\rm GeV} \quad\text{and}\quad 850 \lesssim m_{\mu_D}, m_{\mu^\prime} \lesssim 3000~{\rm GeV}\;.
\]

At larger masses or couplings, the $a_\mu$ contribution decouples and cosmological constraints exclude the parameter space.  
Both BP1 and BP3 lie squarely in this near-resonant band, with $m_{H_D}$ tuned relative to $m_{V_D}$ so that the scalar-mediated annihilation channel efficiently reduces the relic abundance while ensuring the correct sign and magnitude of the $(g-2)_\mu$ shift.

\begin{figure}[h!]
\centering
\begin{subfigure}[b]{0.495\textwidth}
\includegraphics[width=\textwidth]{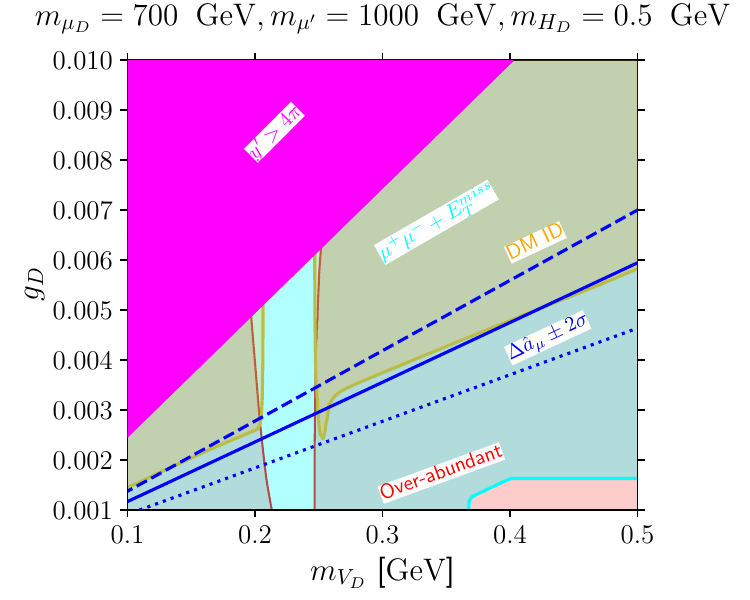}
\centering\small{(a)}
\end{subfigure}
\begin{subfigure}[b]{0.495\textwidth}
\includegraphics[width=\textwidth]{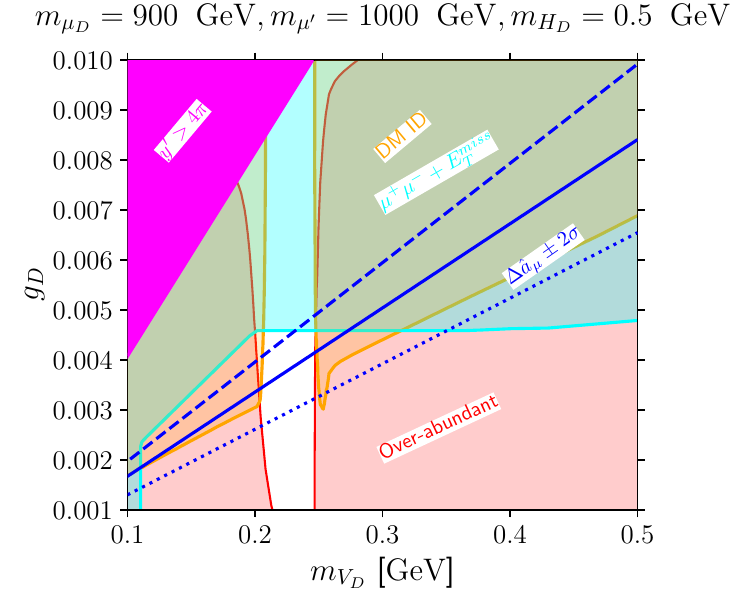}
\centering\small{(b)}
\end{subfigure}\\
\begin{subfigure}[b]{0.495\textwidth}
\includegraphics[width=\textwidth]{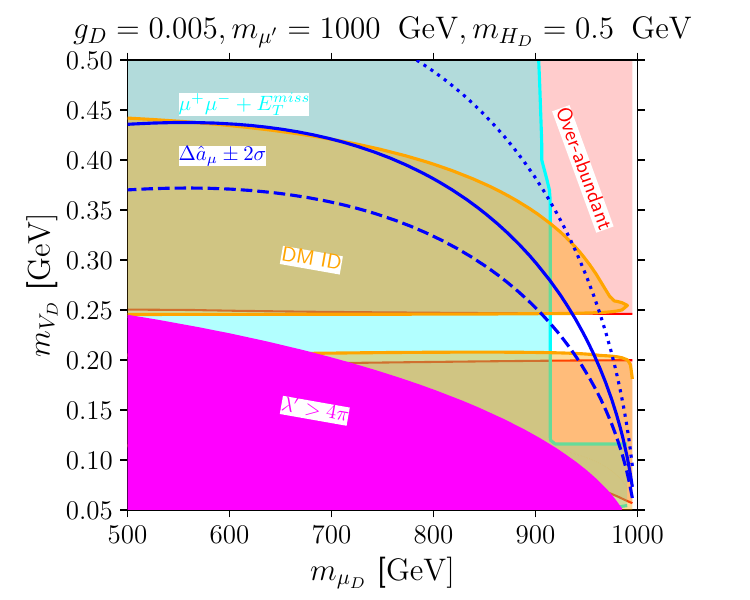}
\centering\small{(c)}
\end{subfigure}
\begin{subfigure}[b]{0.495\textwidth}
\includegraphics[width=\textwidth]{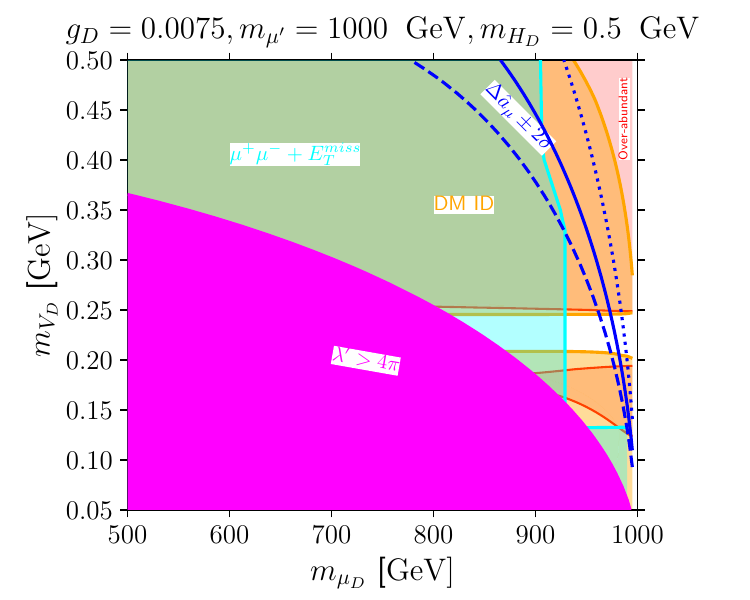}
\centering\small{(d)}
\end{subfigure}
\caption{\label{fig: 2D plots}
The 2D parameter space of $(m_{V_D}, g_D)$ plane with $m_{\mu_D}=700$~GeV (a)
and $900$~GeV (b), with $m_{\mu^\prime}=1000$~GeV and $m_{H_D}=0.5$~GeV,
as well as the $(m_{\mu_D}, m_{V_D})$ plane with $g_D=0.005$ (c) and $0.0075$ (d) 
for the same $m_{\mu^\prime}$ and $m_{H_D}$.  
The magenta, orange, red, and cyan regions are excluded by perturbativity, DM indirect detection, relic density, and collider constraints, respectively.  
The solid red line corresponds to the relic density $\Omega_{\mathrm{DM}}h^2 = 0.12$, the solid orange line indicates the $P_{\mathrm{ann}}=3.2\times10^{-28}$ cm$^3$s$^{-1}$GeV$^{-1}$ limit, and the dotted, solid, and dashed blue lines represent $\Delta\hat{a}_\mu=-2\sigma$, $\Delta\hat{a}_\mu=0$, and $\Delta\hat{a}_\mu=+2\sigma$, respectively, within the {\it tension scenario}.  
The solid cyan line shows the LHC exclusion limit at 95 percent C.L.}
\end{figure}

In \Cref{fig: 2D plots} we present representative projections on the $(m_{V_D}, g_D)$ and $(m_{\mu_D}, m_{V_D})$ planes for different fixed values of $m_{\mu_D}$ and $g_D$.
The magenta, orange, red, and cyan regions are excluded by perturbativity, DM indirect detection (ID), relic density, and collider constraints, respectively.  
The solid red line corresponds to the relic density $\Omega_{\mathrm{DM}}h^2 = 0.12$, while the solid orange line indicates the $P_{\mathrm{ann}} = 3.2\times10^{-28}$ cm$^3$s$^{-1}$GeV$^{-1}$ limit from Planck.  
The dotted, solid, and dashed blue lines represent $\Delta\hat{a}_\mu=-2\sigma$, $\Delta\hat{a}_\mu=0$, and $\Delta\hat{a}_mu=+2\sigma$, respectively, within the {\it tension scenario}.  
The solid cyan line marks the LHC exclusion at 95 percent C.L.

In these panels, the allowed region consistent with all constraints resides in the white area bounded by the blue lines.  
For example, in \Cref{fig: 2D plots}(a) the model is completely excluded, while in \Cref{fig: 2D plots}(b) the viable region corresponds to $0.21\lesssim m_{V_D} \lesssim 0.25$~GeV and $0.0025\lesssim g_D \lesssim 0.0045$.  
Increasing $g_D$ enhances $\Delta a_\mu$, while increasing $m_{V_D}$ suppresses it, in agreement with the analytical scaling $a_\mu \propto g_D^2/m_{V_D}^2$.  
As the mass splitting between $m_{\mu_D}$ and $m_{\mu^\prime}$ decreases, the slope of the blue contours becomes steeper, as seen in \Cref{fig: 2D plots}(a) versus \Cref{fig: 2D plots}(b), reflecting the sensitivity of the loop functions to the mass ratio $r_D$.  
The $(m_{\mu_D}, m_{V_D})$ projections shown in \Cref{fig: 2D plots}(c),(d) further demonstrate how the intersection of relic density, $a_\mu$, and collider constraints isolates a narrow band near the $H_D$ resonance.

In \Cref{fig: 2D plots}, the masses of $\mu^\prime$ and $H_D$ are fixed to common reference values. 
While the former plays only a marginal role -- its mass simply needs to be large enough for $\mu_D$ to exceed the current lower limit of about 850~GeV -- the latter has a more significant impact. 
To illustrate this, consider \Cref{fig: 2D plots}(b): the viable region forms a narrow vertical band in $m_{V_D}$, centred around $0.21$--$0.25$~GeV, where the near-resonant condition $2m_{\mathrm{DM}}\!\lesssim\!m_{H_D}$ ensures consistency with CMB limits. 
Increasing $m_{H_D}$ shifts this band towards larger $m_{V_D}$ values, with its right edge following the relation $m_{V_D}\!\simeq\!m_{H_D}/2$. 
Since the collider and $\Delta\hat a_\mu$ contours are essentially unaffected by $m_{H_D}$, the region of overlap between the band and the allowed $(g-2)_\mu$ area gradually moves into the collider-excluded region (cyan area). 
Conversely, decreasing $m_{H_D}$ pushes the band towards smaller $m_{V_D}$ values, which again fall under the collider bound. 
Thus, the qualitative structure of the allowed parameter space is determined primarily by the resonance condition, rather than by the precise values of $m_{H_D}$ or $m_{\mu^\prime}$.
Benchmark BP1 offers a concrete numerical example of a point lying in such a narrow resonant strip and simultaneously satisfying the $a_\mu$ and relic-density constraints.

To illustrate this interplay further, it is useful to consider how the heavy-lepton sector restricts the shape of the viable domain. 
Both $\mu_D$ and $\mu'$ enter the $(g-2)_\mu$ loop diagrams, yet collider bounds force them above roughly 850 GeV, with the scan favouring values well above 1 TeV. 
This decoupling would ordinarily suppress the loop contribution, but the very light dark photon compensates for this by keeping the relevant mass ratios in the loop functions within the enhanced regime. 
Thus, viable points are driven toward a regime where $m_{\mu_D}$ and $m_{\mu'}$ are heavy enough to satisfy collider bounds but not so heavy as to eliminate their contribution to $(g-2)_\mu$. 
Benchmark BP1 exemplifies this region: a configuration with heavy vector-like leptons still capable of generating the correct magnitude of $\Delta a_\mu$ due to the presence of sub-GeV dark vectors and the resonant condition $m_{H_D}\simeq 2m_{V_D}$.

\subsection{Compatibility scenario}
\label{sec:combined_compatibility}

\begin{figure}[htbp]
\centering
\includegraphics[trim={0 6.15cm 0 0},clip,width=0.8\textwidth]{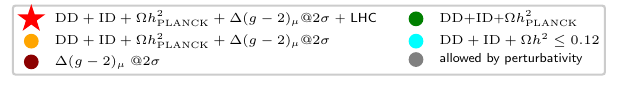}\\
\begin{subfigure}[b]{0.495\textwidth}
   \includegraphics[trim={0 0 0 1.5cm},clip,width=\textwidth]{fig/cosmo+g2+lhc_MVP_GD_25.pdf}\subcaption{}
\end{subfigure}
\begin{subfigure}[b]{0.495\textwidth}
   \includegraphics[trim={0 0 0 1.5cm},clip,width=\textwidth]{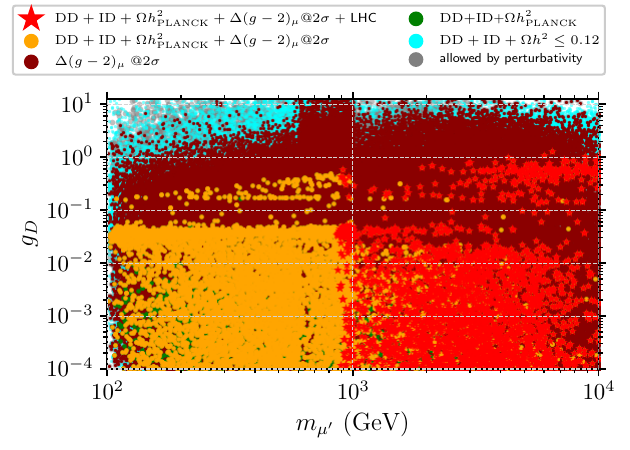}\subcaption{}
\end{subfigure}\\
\begin{subfigure}[b]{0.495\textwidth}
   \includegraphics[trim={0 0 0 1.5cm},clip,width=\textwidth]{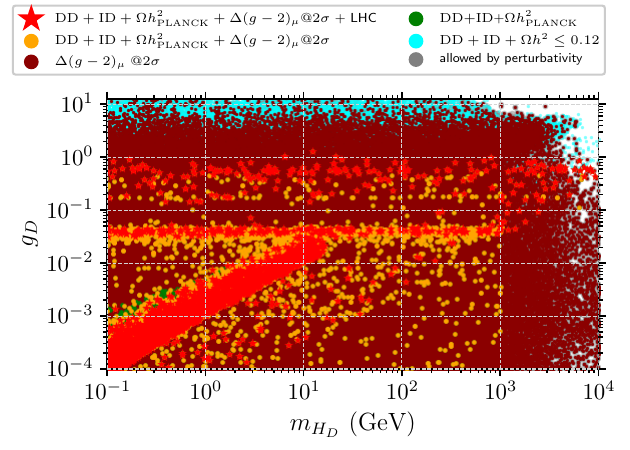}\subcaption{}
\end{subfigure}
\begin{subfigure}[b]{0.495\textwidth}
   \includegraphics[trim={0 0 0 1.5cm},clip,width=\textwidth]{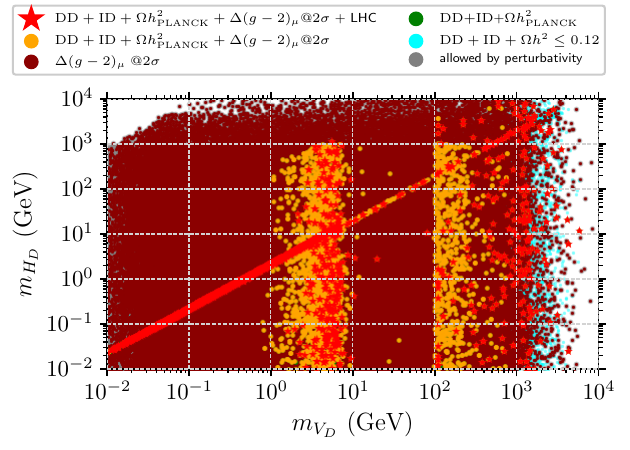}\subcaption{}
\end{subfigure}\\
\begin{subfigure}[b]{0.495\textwidth}
   \includegraphics[trim={0 0 0 1.5cm},clip,width=\textwidth]{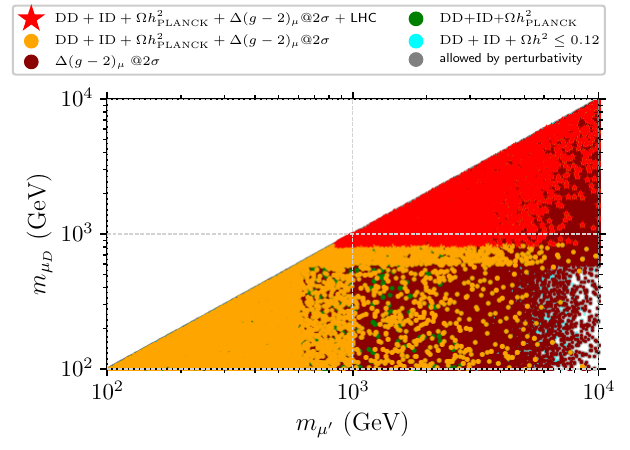}\subcaption{}
\end{subfigure}
\begin{subfigure}[b]{0.495\textwidth}
   \includegraphics[trim={0 0 0 1.5cm},clip,width=\textwidth]{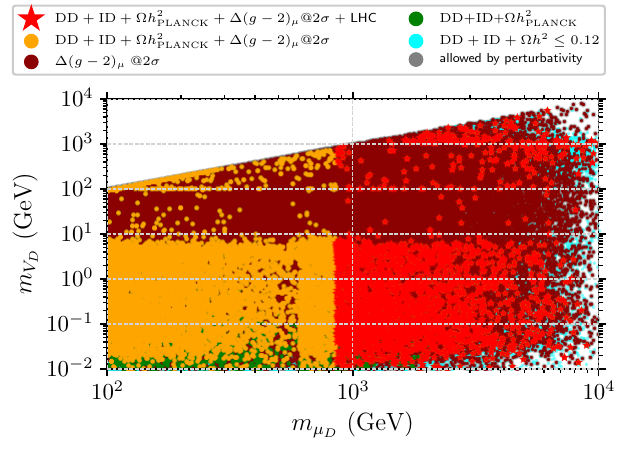}\subcaption{}
\end{subfigure}
\caption{\label{fig:scatter-cosmo-g2-lhc-25}
Scatter plots from the 5D parameter scan projected onto 
(a) $(m_{V_D}, g_D)$, (b) $(m_{\mu_D}, g_D)$, 
(c) $(m_{H_D}, g_D)$, (d) $(m_{H_D}, m_{V_D})$, 
(e) $(m_{\mu^\prime}, m_{\mu_D})$, and (f) $(m_{\mu_D}, m_{V_D})$ planes. 
The red points indicate regions consistent with all constraints, including perturbativity, collider limits, cosmological bounds, and $a_\mu$ (within the {\it compatibility scenario} at the $2\sigma$ level). 
Grey points satisfy only perturbativity. 
Green and cyan points correspond to cosmologically allowed regions (matching or below the Planck relic density). 
Dark-red points satisfy the $a_\mu$ constraint alone, while orange points represent the parameter space consistent with the combined cosmological and $a_\mu$ constraints.}
\end{figure}

In \Cref{fig:scatter-cosmo-g2-lhc-25} we present the analogous results for the {\it compatibility scenario}, in which the measured $a_\mu$ agrees with the SM prediction within uncertainties.  
In this case, the requirement of a large positive $\Delta a_\mu$ is lifted, and the allowed parameter space expands dramatically.  
The same colour code is used as in the previous figure, with red points marking regions satisfying all combined constraints.
The overall colour scheme and structure of the panels are identical to those in \Cref{fig:scatter-cosmo-g2-lhc}, allowing a direct visual comparison of how the parameter space relaxes once the $a_\mu$ tension is lifted.

The most striking feature is that the orange and red regions now occupy a much wider area in all panels, indicating that both light and heavy dark matter masses are allowed.  
The viable $g_D$ range extends from $10^{-4}$ up to order unity, and 
dark vector masses can vary from sub-GeV to several TeV without violating any constraints.  
The resonance condition $2m_{\mathrm{DM}}\simeq m_{H_D}$ still appears as a preferred band, particularly at low masses, where it provides a natural way to evade CMB limits via the temperature-dependent suppression mechanism discussed earlier.  
However, this resonance is no longer the only viable configuration, as non-resonant regions also remain allowed.

The enlarged parameter space in the compatibility scenario can also be interpreted geometrically. 
Once the requirement of generating the full $(g-2)_\mu$ excess is removed, the constraint hypersurfaces widen and decouple from the narrow resonance funnel. 
Instead of sitting on a highly tuned strip in the $(m_{V_D}, g_D)$ plane, compatible points form a two-dimensional band extending towards both smaller coupling values and larger dark photon masses. 
The relic-density constraint alone still shapes the parameter space into resonant and non-resonant regions, but these regions now remain viable without being forced to intersect a specific $(g-2)_\mu$ contour. 
Benchmark BP2, for instance, resides on such a broadened sheet: its parameters are aligned with the relic-density requirement and CMB safety, but its predicted $(g-2)_\mu$ shift lies naturally within the experimentally allowed range without relying on resonance-enhanced loop contributions.

In the $(m_{\mu_D},m_{\mu^\prime})$ and $(m_{\mu_D},m_{V_D})$ panels, the red points extend to multi-TeV vector-like lepton masses, demonstrating that heavy partners decouple smoothly while preserving consistency with all constraints.  
Collider bounds dominate the upper exclusion region, while perturbativity defines the high-coupling cutoff.  
The relic density and direct-detection constraints primarily control the low-mass edge of the parameter space.

The long-lived benchmark BP4 demonstrates an additional feature of the compatibility regime: the possibility of realising dark photons below the dimuon threshold, $m_{V'}<2m_\mu$, without jeopardising cosmological or collider viability. 
In the tension scenario such points are typically disfavoured because the reduced accessibility of the $\mu^+\mu^-$ decay channel weakens the loop contributions to $(g-2)_\mu$. 
However, in the compatibility framework this suppression is acceptable, and the model can instead exploit the modified lifetime and branching ratios of $V'$ to produce distinctive collider signatures such as displaced $e^+e^-$ vertices or invisible-like cascades. 
Once the $(g-2)_\mu$ requirement is removed, there is no longer a need for light $m_{V_D}$ to enhance the loop contribution to $\Delta a_\mu$. 
As a result, the strong correlation between $m_{H_D}$ and $m_{V_D}$ present in the tension scenario relaxes, and the cosmological constraints alone allow the viable region in the $(m_{H_D},m_{V_D})$ plane to extend toward much larger values of $m_{V_D}$.

Benchmark BP2 (short-lived) and BP4 (long-lived), shown in Table~\ref{tab:mergedBPs}, are concrete examples of viable points in this enlarged compatibility region. 
In these benchmarks the coupling $g_D$ and the $H_D$-$V_D$ mass relation are adjusted so that the MPVDM loop contributions to $a_\mu$ fall within the experimentally allowed band without requiring a significant upward shift. 
At the same time, the dark matter relic density remains close to the observed value, and collider constraints are satisfied through appropriately chosen heavy-lepton masses.

\section{MPVDM Benchmarks}
\label{sec:benchmarks}
In this section we describe benchmark points that summarise the key collider, 
cosmological and precision features of the MPVDM model discussed in the previous 
sections. These benchmarks, shown in \Cref{tab:mergedBPs}, incorporate the 
results of the complete 5D parameter scan performed in 
Section~\ref{sec:combined} and were selected to illustrate the characteristic 
phenomenology of both the {\it tension} and {\it compatibility} scenarios for 
$(g-2)_\mu$.

BP1 and BP3 capture the tension scenario in its short- and long-lived 
incarnations, highlighting how the scalar resonance and the light $V'$ 
cooperate to reproduce a large positive contribution to $\Delta a_\mu$ while 
simultaneously satisfying stringent collider and cosmological constraints. 
BP2 and BP4 illustrate the broader flexibility allowed in the compatibility 
scenario, in which the $(g-2)_\mu$ measurement no longer forces the model into a 
narrow low-mass corridor, thus permitting heavier dark-matter candidates, 
different kinematical configurations, and long-lived dark-sector signatures. 
By mapping these explicit points onto the scatter plots and analytical 
considerations of Section~\ref{sec:combined}, one sees that the viable parameter 
space is both structured and predictive, offering concrete targets for HL-LHC 
searches and future precision probes of light dark sectors. 
They also demonstrate how the collider signatures evolve depending on the 
lifetime of $V^\prime$, thereby covering the full range of possibilities 
relevant for present and future LHC searches.

The left block of \Cref{tab:mergedBPs} corresponds to the {\it short-lived 
regime}, in which $V'$ decays promptly (BP1) or nearly promptly (BP2), producing 
high-multiplicity muon final states. BP1 represents the short-lived version of 
the {\it tension scenario}, while BP2 illustrates the compatibility scenario in 
which heavier dark-matter masses (here $m_{V_D}=300$ GeV) are allowed. This 
difference in the dark-sector scales leads to a noteworthy collider consequence: 
in BP1 the light dark sector produces very collimated, boosted muon pairs from 
$V'\to\mu^+\mu^-$ (as also discussed in \Cref{sec:multileptons}), whereas in BP2 
the heavier kinematics reduce the boost and yield wider muon separations. 
This reduced collimation improves experimental muon identification and reconstruction efficiency compared to BP1, where dedicated lepton-jet–like strategies may be required.

The right block of the table corresponds to the {\it long-lived regime}.  
Here \(V'\) is sufficiently light that the dimuon channel is kinematically forbidden and its dominant decay modes become:
\[
V' \to e^+ e^- \quad ( \sim 31\% ), 
\qquad 
V'\to \nu\nu \quad (\sim 60\%).
\]
As a consequence, BP3 and BP4 no longer generate prompt or collimated multi-muon topologies.  
Instead, they yield displaced pairs of electrons, with or without missing transverse energy, depending on whether the decay proceeds via \(e^+e^-\) or neutrino pairs.  
Moreover, since only a fraction of \(V'\) decays visibly into electrons, signatures with four visible electrons (originating from two \(V'\) decays) are intrinsically rarer than in the short-lived muon-dominated case.  
These long-lived benchmarks therefore motivate displaced-vertex searches targeting \(e^+e^-\) pairs with characteristic radial distributions and missing energy from the neutrino channels.

\begin{table*}[htb]
\setlength{\tabcolsep}{10pt} 
\centering
\scriptsize
\renewcommand{\arraystretch}{1.1}
\hspace*{-22pt}\begin{tabular}{|l|c|c||l|c|c|}
\toprule
\multicolumn{3}{|c||}{\textbf{Short-lived scenario}} &
\multicolumn{3}{c|}{\textbf{Long-lived scenario}} \\
\midrule
Inputs / Observables 
 & BP1 & BP2 & Inputs / Observables & BP3 & BP4 \\
 & ({\it tension}) & ({\it compatibility}) & & ({\it tension}) & ({\it compatibility}) \\
\midrule
$g_D$            & 0.003 & 0.13 & 
$g_D$            & 0.0023 & 0.002 \\
$m_{V_D}$ [GeV]   & 0.28  & 300 & 
$m_{V_D}$ [GeV]   & 0.20 & 0.20 \\
$m_{\mu_D}$ [GeV]   & 900 & 1100 & 
$m_{\mu_D}$ [GeV]  & 1000 & 1000 \\
$m_{\mu^\prime}$ [GeV] & 1000 & 1200 & 
$m_{\mu^\prime}$ [GeV] & 1200 & 1100 \\
$m_{H_D}$ [GeV] & 0.677 & 0.677 & 
$m_{H_D}$ [GeV] & 0.485 & 0.48 \\
$m_{V^\prime}$ [GeV] & 0.276 & 300 & 
$m_{V^\prime}$ [GeV] & 0.196 & 0.199 \\
\midrule
$\Delta \hat{a}_\mu^{t,c}$
                   & $-1.97$ & $-0.600$ & 
$\Delta \hat{a}_\mu^{t,c}$ 
                  & $-1.63$ &$0.685$\\
$\Omega_{\rm DM}h^2$ 
                 & 0.118 & 0.110 & 
$\Omega_{\rm DM}h^2$ 
                 & 0.106 & 0.0936 \\
$N_{\rm events}$ & 
                $9.46\times10^{-8}$ & $1.69$ &
$N_{\rm events}$ & 
                $5.13\times10^{-9}$ & $8.93\times10^{-10}$ \\
$\hat{p}_{\rm DD}$ & $\simeq 1$ & $0.185$ & 
$\hat{p}_{\rm DD}$ & $\simeq 1$ & $\simeq 1$ \\
$\hat{\sigma}_{\rm ann}^{\rm ID}$
                        & 0.964 & $7.85\times 10^{-5}$ & 
$\hat{\sigma}_{\rm ann}^{\rm ID}$ 
                        & 0.884 & 0.245 \\
\midrule
$\tau_{V^\prime}$ [ns] 
       & $1.31\times10^{-6}$ & $1.50\times10^{-12}$ & 
$\tau_{V^\prime}$ [ns] 
      & $2.60\times10^{-3}$ & $1.92\times10^{-2}$ \\
$\ell_{V^\prime}$ [$\mu$m] 
               & $0.39\gamma$ & $4.5\time 10^{-7}\gamma$&
$\ell_{V^\prime}$ [$\mu$m]
               & $1430\gamma$ & $5760\gamma$ \\
$\epsilon_{AV^\prime}$ 
& $1.05\times10^{-5}$ & $2.27\times10^{-4}$ & $\epsilon_{AV^\prime}$ 
& $7.57\times10^{-6}$ & $3.74\times10^{-6}$ \\
\midrule
$Br(\mu^\prime \to V^\prime\mu)$  & 0.401 & 0.659 & 
$Br(\mu^\prime \to V^\prime\mu)$  & 0.416 & 0.470 \\
$Br(\mu^\prime \to H_D\mu)$  & 0.388 & 0.341 & 
$Br(\mu^\prime \to H_D\mu)$  & 0.401 & 0.467 \\
$Br(\mu^\prime \to V_D\mu_D)$ & 0.211 & 0 & 
$Br(\mu^\prime \to V_D\mu_D)$ & 0.183 & 0.062 \\
\midrule
$Br(H_D \to V_DV_D^*)$        & 0.640 & 0.667 & 
$Br(H_D \to V_DV_D^*)$        & 0.634 & 0.659 \\
$Br(H_D \to V^\prime V^\prime)$ & 0.353 & 0.333 & 
$Br(H_D \to V^\prime V^\prime)$ & 0.356 & 0.337 \\
$Br(H_D \to \mu^+\mu^-)$ 
     & $7.8\times10^{-3}$ &$3.81\times10^{-9}$ & 
$Br(H_D \to \mu^+\mu^-)$ 
     & $1.09\times10^{-2}$ & $3.77\times10^{-3}$ \\
\midrule
$\sigma_{\mathrm{tot}}(pp\!\to\!\mu'\mu')$ [fb] 
      & $5.10\times10^{-2}$ & $3.00\times10^{-2}$ &
$\sigma_{\mathrm{tot}}(pp\!\to\!\mu'\mu')$ [fb] 
     & $3.00\times10^{-2}$ & $5.10\times10^{-2}$ \\
\midrule
$Br(\mu'\!\to\!\mu\  V_DV_D)$           & 0.459 & 0.227   & 
$Br(\mu'\!\to\!\mu +V_DV_D/\nu\nu/4\nu)$ & 0.816 & 0.796 \\
$Br(\mu'\!\to\!3\mu)$         & 0.404 & 0.660 & 
$Br(\mu'\!\to\!\mu  e^+e^- )$ & 0.129 & 0.146 \\
$Br(\mu'\!\to\!5\mu\!)$               & 0.137 & 0.131 & 
$Br(\mu'\!\to\!\mu e^+e^-\nu\nu)$ & 0.055 & 0.042 \\
\midrule
$P(\mu'\mu'\!\to\!6\mu (+ \slashed{E}_T)$      & 0.289 & 0.487 & 
$P(\mu'\mu'\!\to\!2\mu 2e + \slashed{E}_T )$ 
&0.300 & 0.299 \\
$P(\mu'\mu'\!\to\!8\mu)$               & 0.111 & 0.150 & 
$P(\mu'\mu'\!\to\!2\mu 4e + \slashed{E}_T)$ 
& $1.72\times 10^{-2}$& $1.41\times 10^{-2}$ \\
$P(\mu'\mu'\!\to\!10\mu)$  & 0.019 & 0.013 & $P(\mu'\mu'\!\to\!2\mu 4e)$ 
                  & $1.66\times 10^{-2}$ & $2.12\times 10^{-2}$ \\
\midrule
$N_{\mathrm{events}}(pp\!\to\!6\mu)$ 
& 44.2 & 43.8 & $N_{\mathrm{events}}(pp\!\to\!2\mu 2e+ \slashed{E}_T)$ 
&27.0 & 45.7 \\
$N_{\mathrm{events}}(pp\!\to\!8\mu)$ 
& 16.9 & 13.5& $N_{\mathrm{events}}(pp\!\to\!2\mu 4e + \slashed{E}_T)$ 
& 1.55 & 2.15\\
$N_{\mathrm{events}}(pp\!\to\!10\mu)$ & 2.9 & 1.2 & $N_{\mathrm{events}}(pp\!\to\!2\mu 4e )$ & 1.50 & 3.25 \\
\bottomrule
\end{tabular}
\caption{\label{tab:mergedBPs}
Benchmarks for the MPVDM model.
The left block (BP1--BP2) corresponds to short-lived scenarios yielding prompt multi-muon signatures, while the right block (BP3--BP4) includes heavier or long-lived $V'$
states.
Each block contains two cases: the {\it tension} scenario, consistent with a positive $(g-2)_\mu$ deviation, and the {\it compatibility} scenario, where experiment and theory agree within uncertainties.
All points satisfy cosmological and collider limits.
In case of   long-lived scenario $V'$
decays about $31\%$ to $e^+e^-$
and $60\%$ to neutrinos.
For all benchmarks $\mu_D$ decays 100\% to $\mu V_D$.
Event yields are computed for $\mathcal{L}=3000$~fb$^{-1}$ at $\sqrt{s}=14$~TeV (HL-LHC).
}
\end{table*}
The branching ratios listed for $\mu^\prime$, $H_D$ and $V^\prime$ encode the structure of the cascade decays responsible for all these multi-lepton signatures.  
For the short-lived regime (BP1, BP2), the inclusive probabilities $P(\mu^\prime\mu^\prime \to n\mu)$ quantify the likelihood of producing six, eight or ten muons from pair-produced heavy leptons.  
These probabilities remain sizeable even in the compatibility benchmark BP2, reflecting the stability of the short-lived phenomenology across different mass scales.
In contrast, the long-lived benchmarks (BP3, BP4) no longer yield multi-muon final states.  
Instead, the relevant probabilities describe the inclusive rates for the mixed muon–electron signatures 
\(
2\mu + 2e + \slashed{E}_T,\;
2\mu + 4e + \slashed{E}_T,\;
2\mu + 4e,
\)
whose relative sizes depend on the visible versus invisible branching fractions of the long-lived \(V'\).
Here, the missing transverse energy $\slashed{E}_T$ arises from neutrinos and/or dark-sector particles produced in the cascade decays of $V'$, $H_D$, or $\mu'$.
 
The resulting event yields, shown in the bottom rows of Table~\ref{tab:mergedBPs}, demonstrate that even displaced-electron signatures can reach observable rates at the HL-LHC.

Finally, the table includes predictions for the total production cross sections $\sigma(pp\to\mu^\prime\mu^\prime)$ and the corresponding HL-LHC event counts at $\sqrt{s}=14$~TeV and $\mathcal{L}=3000\,{\rm fb}^{-1}$.  
These benchmarks collectively serve as concrete, high-level summaries of the model’s phenomenology, providing representative collider targets across both prompt and displaced regimes.  
They can therefore be used directly as reference points for collider reinterpretation studies, detector-performance evaluations, and the design of new searches for light dark sectors coupled to vector-like leptons.


\section{Conclusions}
\label{sec:conclusions}

In this study, we have investigated the Muonic Portal to Vector Dark Matter (MPVDM) model, a theoretically well-motivated framework that links the origin of dark matter to the muon sector via new vector-like leptons (a $\Z_2$-odd $\mu_D$ and a $\Z_2$-even $\mu^\prime$) and a dark $\Z_2$-even scalar mediator ($H_D$). 
This setup provides a unified mechanism capable of simultaneously addressing the long-standing muon anomalous magnetic moment ($a_\mu$) anomaly (either assuming it is still present or not) and the cosmological dark matter abundance. 

We derived and validated the full expressions for the MPVDM contributions to $a_\mu$, explicitly checking consistency with existing results and extending the analysis to include both the {\it tension} and {\it compatibility} scenarios that reflect the evolving experimental--theoretical status of $(g-2)_\mu$. 
In the {\it tension scenario}, where a positive deviation $\Delta a_\mu \sim 2.5\times10^{-9}$ is required, the model predicts a sharply defined and highly constrained region of parameter space. 
To generate a sufficiently large $\Delta a_\mu$ despite the heavy vector-like leptons, dark matter must be light ($m_{\mathrm{DM}}\lesssim1$~GeV) and coupled with a small gauge coupling $g_D\sim10^{-3}$. 
Such light dark matter would normally be excluded by CMB bounds on s-wave annihilation, but in the MPVDM it naturally resides near the scalar resonance, $2m_{\mathrm{DM}}\simeq m_{H_D}$, where the annihilation cross section exhibits a {\it generic off-resonance velocity-suppression mechanism} -- one of the new key findings of this study. 
This kinematical effect, realised when $2m_{\mathrm{DM}}$ lies below about 10--20\% of the resonance, ensures that the relic abundance remains consistent with the Planck limit while strongly suppressing late-time annihilation, allowing light dark matter to remain viable without invoking the ultra-narrow Breit--Wigner resonances and associated early kinetic decoupling assumed in earlier studies. 
Importantly, this behaviour does not require fine-tuning of parameters and is expected to occur generically in near-resonant thermal dark matter scenarios. 
At the same time, the light dark matter mass enhances the loop contribution to $a_\mu$, compensating for the suppression from the heavy vector-like states. 
This interplay between cosmology and particle physics renders the {\it tension scenario} highly predictive and phenomenologically testable.

In contrast, the {\it compatibility scenario}, consistent with the 2025 Muon $g-2$ Theory Initiative update, exhibits a much broader viable parameter space. 
Here, no large positive $\Delta a_\mu$ is required, allowing both resonant and non-resonant configurations across a wide mass range. 
Dark vector masses can span from sub-GeV to multi-TeV, with couplings varying from $10^{-4}$ to ${\cal O}(1)$, and the relic density constraint can be satisfied naturally without fine-tuning. 
The same off-resonance suppression mechanism remains operative in the low-mass regime, ensuring that the model smoothly transitions between the two scenarios as experimental inputs evolve. 
By slightly reducing $g_D$ and increasing $m_{H_D}$, the benchmark points that explain the $a_\mu$ tension can also satisfy the {\it compatibility scenario}, highlighting the model's robustness and flexibility. 
We have visualised these combined constraints through five-dimensional scatter plots and two-dimensional parameter projections, which clearly identify the viable regions and their correlations among the model parameters.

We have also performed a comprehensive collider analysis by recasting ATLAS and CMS searches for dilepton plus missing transverse energy signatures, setting a lower bound of approximately 850~GeV on the masses of the $\Z_2$-odd vector-like muons, almost independently of the mass of the $\Z_2$-even one. 
The parameter space allowed by all the constraints predicts striking multi-lepton signatures with up to six, eight, or ten muons in the final state, accompanied by missing energy from the dark sector -- a distinctive experimental hallmark for future searches at the LHC in its high-luminosity phase and at future colliders.

To make these predictions concrete, we have identified a set of representative 
benchmark points that realise the different regimes uncovered in our analysis -- 
including prompt multi-muon signatures and displaced or mixed electron--muon 
topologies in the long-lived case. These benchmarks provide realistic collider 
targets for the HL-LHC and offer clear guidance for future dedicated searches. 
Importantly, they also show that the predicted signal rates remain sizeable even 
when the vector-like muon partner $\mu'$ lies above the TeV scale, indicating 
that distinctive MPVDM signatures can persist well into the high-mass regime and 
thus merit detailed experimental investigation.

The MPVDM scenario thus provides a coherent, predictive, and phenomenologically rich framework connecting the muon sector to the dark sector. 
Its ability to accommodate both the presence and absence of the $(g-2)_\mu$ anomaly, while naturally satisfying cosmological and collider bounds, makes it a compelling target for future experimental investigation. 
Further exploration of displaced dimuon and displaced electron-pair signatures from long-lived dark photons, along with precision measurements of $a_\mu$, will be crucial in probing this model. 
In summary, the MPVDM framework offers a unified and testable explanation for the interplay between dark matter and muon physics, bridging cosmological observations and collider phenomenology, and revealing a generic near-resonant mechanism through which light (sub- to few-GeV) thermal dark matter can remain consistent with all current data.


\section*{Acknowledgements}
The authors would like to thank Prof. Douglas Ross for his valuable discussions, insightful comments, and guidance in the loop calculations of the MPVDM  contribution to $a_\mu$.
Authors acknowledge the use of the IRIDIS High-Performance Computing Facility and associated support services at the University of Southampton in completing this work.
AB is supported in part through the
NExT Institute and  STFC Consolidated Grant No. ST/L000296/1.
AB acknowledges support from the Leverhulme Trust project MONDMag (RPG-2022-57).
LP's work is supported by ICSC – Centro Nazionale di Ricerca in High Performance Computing, Big Data and Quantum Computing, funded by European Union – NextGenerationEU.
NT has received funding support from the NSRF via the Program
Management Unit for Human Resources \& Institutional Development, Research and
Innovation [grant number B13F670063].
FW would like to thank Prof. Rikard Enberg for his helpful comments and suggestions.

\clearpage
\appendix

\section{Analytical results for $a_\mu$ in the MPVDM}
\label{app:g-2analytical}

In general, the Standard Model (SM) prediction for the anomalous magnetic moment of the muon can be separated into three sectors: 
(1)~the pure Quantum Electrodynamics (QED) contribution, 
(2)~the Electroweak (EW) contribution, and 
(3)~the Hadronic Vacuum Polarisation (HVP) contribution.%
\footnote{We adopt the SM prediction based on the 2025 Muon $g\!-\!2$ Theory Initiative update~\cite{Aliberti:2025WP}, which uses a lattice-informed HVP evaluation. 
Independent lattice calculations from the BMW collaboration~\cite{Borsanyi:2020mff} and from the Fermilab/HPQCD/MILC collaboration~\cite{Bazavov:2024PRL} are consistent within uncertainties, lending confidence to the lattice-based result. 
The earlier data-driven dispersive evaluations yield a smaller HVP contribution and therefore a lower total SM prediction.}

The theoretical uncertainties from the QED and EW sectors are negligibly small, while the hadronic contribution remains the dominant source of uncertainty in the SM prediction. 
As discussed in~\Cref{sec:intro}, the comparison between the latest experimental world average~\cite{Muong-2:2025xyk} and the updated SM prediction~\cite{Aliberti:2025WP} shows no statistically significant deviation within present uncertainties. 
However, if one instead adopts the pre-2023 data-driven evaluations of the HVP, the well-known excess 
$\Delta a_\mu^{\rm EXP}\!\approx\!(2.5\pm0.5)\!\times\!10^{-9}$ (corresponding to about $5\sigma$) reappears. 
Both interpretations are therefore of current interest.
 In the following, we present the predictions for the contributions to $a_\mu$ within our model. 
Throughout this analysis we neglect diagrams involving scalar mixing, {\it i.e.} we assume $\sin\theta_S = 0$.

\begin{figure}[thb]
\centering
    \begin{subfigure}[b]{0.32\textwidth}
		\centering	
        \includegraphics[width=\textwidth]{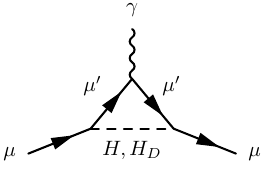}
        \caption{\label{fig:feyngm2a}}
	\end{subfigure}\hfill
    \begin{subfigure}[b]{0.32\textwidth}
	      \centering	
        \includegraphics[width=\textwidth]{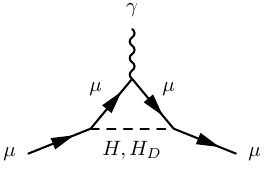}
        \caption{\label{fig:feyngm2b}}
	\end{subfigure}\hfill
    \begin{subfigure}[b]{0.32\textwidth}
		\centering	
        \includegraphics[width=\textwidth]{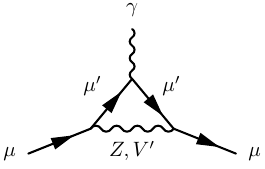}
        \caption{\label{fig:feyngm2c}}
	\end{subfigure}\\
    \begin{subfigure}[b]{0.32\textwidth}
		\centering
        \includegraphics[width=\textwidth]{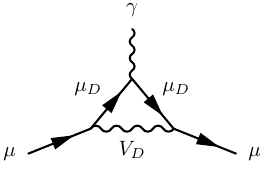}
        \caption{\label{fig:feyngm2d}}
	\end{subfigure}\hfill
    \begin{subfigure}[b]{0.32\textwidth}
		\centering	
        \includegraphics[width=\textwidth]{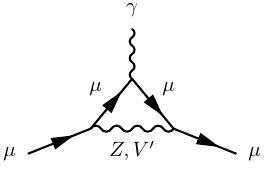}
        \caption{\label{fig:feyngm2e}}
	\end{subfigure}\hfill
    \begin{subfigure}[b]{0.32\textwidth}
		\centering	
        \includegraphics[width=\textwidth]{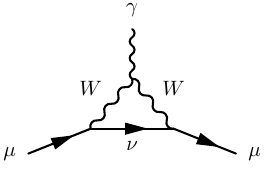}
        \caption{\label{fig:feyngm2f}}
	\end{subfigure}
\caption{Diagrams contributing to $a_\mu$ in the MPVDM. Those involving only SM particles provide a new physics contribution through the muon mixing angles.}
\label{fig:feyngm2}
\end{figure}

The loop diagrams from MPVDM at the leading order are depicted in~\Cref{fig:feyngm2}. 
The general formulae to compute the contribution of NP at leading order are well known from established literature~\cite{Leveille:1977rc,Moore:1984eg}, which we report here for completeness.

The contribution from diagrams involving a neutral scalar $S$ and a muon or muon partner $f$ is given by:
\begin{equation}
a_\mu^S={m_\mu^2\over 8\pi} \int_0^1 dx {C_S^2\left(x^2\left(1+{m_f\over m_\mu}\right)-x^3\right)+C_P^2\left(m_f\to-m_f\right)\over m_\mu^2 x^2+(m_f^2-m_\mu^2)x+m_S^2(1-x)}\;,
\end{equation}
where $C_S$ and $C_P$ are the scalar and pseudo-scalar couplings in the scalar-fermion-fermion vertices and $m_S$ represents the scalar mass ($H$ or $H_D$).

If $f=\mu^\prime$ (\Cref{fig:feyngm2a}) and  $m_\mu\ll m_{\mu^\prime}$, the integration leads to:
\begin{equation}
a_\mu^{(a)}[S\mu^\prime] = \frac{(C_S^2 - C_P^2)}{16\pi^2}\frac{m_\mu m_{\mu^\prime}}{(m_{\mu^\prime}^2 - m_S^2)^3} \left( m_{\mu^\prime}^4 -4m_{\mu^\prime}^2 m_S^2 + 3 m_S^4 -2 m_S^4 \log{\left(\frac{m_S^2}{m_{\mu^\prime}^2}\right)} \right)+ \mathcal{O}\left(\frac{m_\mu^2}{m_{\mu'}^2}\right)\;,  
\end{equation}
while when $f=\mu$ (\Cref{fig:feyngm2b}), and in two opposite limits, the integration simplifies to:
\begin{eqnarray}
m_\mu \ll m_S: &&\quad a_\mu^{(b)}[S\mu]= {m_\mu^2 \over 48\pi^2 m_S^2}\left( (11C_P^2 - 7 C_S^2) - 12(C_S^2 - C_P^2)\log{\left(\frac{m_\mu}{m_S}\right)} \right) 
+ \mathcal{O}\left(\frac{m_\mu^4}{m_{S}^4}\right)
\\
m_S \ll m_\mu: &&\quad a_\mu^{(b)}[S\mu] = \frac{3C_S^2-C_P^2}{16\pi^2}  + \mathcal{O}\left(\frac{m_\mu^2}{m_{S}^2}\right)
\end{eqnarray}

By substituting the expressions of the couplings from~\Cref{tab:couplings of VVA/VVZ vertex for MPVDM}, we  obtain the contributions from all particles in  loops for MPVDM. 

\begin{table}[htbp]
		\setlength{\tabcolsep}{15pt} 
		\renewcommand{\arraystretch}{2} 
		\centering
            \resizebox{\columnwidth}{!}{%
		\begin{tabular}{ |c|c|c| } 
			\hline
	            \bf{SFF-vertices} & \bf{Scalar couplings} $(C_S)$
 & \bf{Psuedo-scalar couplings} $(C_P)$ \\
            \hline
			$H \mu^+ \mu^-$ & $\displaystyle -\frac{y}{\sqrt{2}}\cos{\theta}_L\cos{\theta}_R$ & $0$ \\
			$H_D \mu^+ \mu^-$ & $\displaystyle \frac{y^\prime}{\sqrt{2}}\sin{\theta}_L\cos{\theta}_R$ & $0$ \\
			$H \mu^{\prime +} \mu^{\prime -}$ & $\displaystyle -\frac{y}{2\sqrt{2}}(\sin{\theta}_R\cos{\theta}_L + \sin{\theta}_L\cos{\theta}_R)$ & $\displaystyle -\frac{y}{2\sqrt{2}}(\sin{\theta}_R\cos{\theta}_L - \sin{\theta}_L\cos{\theta}_R)$ \\
			$H_D \mu^{\prime +} \mu^{\prime -}$ & $\displaystyle -\frac{y}{2\sqrt{2}}(\cos{\theta}_L\cos{\theta}_R - \sin{\theta}_L\sin{\theta}_R)$ & $\displaystyle -\frac{y}{2\sqrt{2}}(\cos{\theta}_L\cos{\theta}_R + \sin{\theta}_L\sin{\theta}_R)$ \\
			$H \mu^{\prime +} \mu^-$ & $\displaystyle -\frac{y}{2\sqrt{2}}(\sin{\theta_R}\cos{\theta_L}+\sin{\theta_L}\cos{\theta_R})$ & $\displaystyle \frac{y}{2\sqrt{2}}(\sin{\theta_R}\cos{\theta_L}-\sin{\theta_L}\cos{\theta_R})$ \\
			$H_D \mu^{\prime +} \mu^-$ & $\displaystyle -\frac{y^\prime}{2\sqrt{2}}(\cos{\theta_L}\cos{\theta_R}-\sin{\theta_L}\sin{\theta_R})$ & $\displaystyle \frac{y^\prime}{2\sqrt{2}}(\sin{\theta_L}\sin{\theta_R}+\cos{\theta_L}\cos{\theta_R})$ \\
			\hline		\textbf{VFF- vertices} & \textbf{Vector couplings $(C_V)$} & \textbf{Axial couplings $(C_A)$}\\
			\hline
			$\gamma \mu^+ \mu^-$ & $-e$ & $0$ \\ 
			$\gamma \mu^+_D \mu^-_D$ & $-e$ & $0$ \\
			$\gamma \mu^{\prime +} \mu^{\prime -}$ & $-e$ & $0$ \\
			$Z \mu^+ \mu^-$ & $\displaystyle -\frac{g_W}{c_W}\left(\frac{\cos^2{\theta}_L}{4} - s_W^2\right)$ & $\displaystyle -\frac{g_W}{c_W}\frac{\cos^2{\theta}_L}{4}$\\ 
			$Z \mu^{\prime +} \mu^-$ & $\displaystyle -\frac{g_W}{c_W}\frac{\sin{\theta}_L\cos{\theta}_L}{4}$ & $\displaystyle -\frac{g_W}{c_W}\frac{\sin{\theta}_L\cos{\theta}_L}{4}$ \\
			$Z \mu^{\prime +} \mu^{\prime -}$ & $\displaystyle -\frac{g_W}{c_W}\left(\frac{\sin^2{\theta_L}}{4}-s_W^2\right)$ & $\displaystyle -\frac{g_W}{c_W}\frac{\sin^2{\theta_L}}{4}$ \\
			$Z \mu_D^+ \mu_D^-$ & $\displaystyle \frac{g_W s_W^2}{c_W}$ & $0$ \\
			$V_0 \mu^+ \mu^-$ & $\displaystyle -\frac{g_D}{4}(\sin^2{\theta}_L + \sin^2{\theta}_R)$ & $\displaystyle -\frac{g_D}{4}(\sin^2{\theta}_L - \sin^2{\theta}_R)$ \\
			$V_0 \mu^{\prime +} \mu^-$ & $\displaystyle \frac{g_D}{4}(\sin{\theta}_L\cos{\theta}_L + \sin{\theta}_R\cos{\theta}_R)$ & $\displaystyle \frac{g_D}{4}(\sin{\theta}_L\cos{\theta}_L - \sin{\theta}_R\cos{\theta}_R)$ \\
			$V_D \mu^+_D \mu^-$ & $\displaystyle -\frac{g_D}{2\sqrt{2}}(\sin{\theta}_L+\sin{\theta}_R)$ & $\displaystyle -\frac{g_D}{2\sqrt{2}}(\sin{\theta}_L-\sin{\theta}_R)$ \\
			$V_D \mu^+_D \mu^{\prime -}$ & $\displaystyle \frac{g_D}{2\sqrt{2}}(\cos{\theta}_L+\cos{\theta}_R)$ & $\displaystyle \frac{g_D}{2\sqrt{2}}(\cos{\theta}_L-\cos{\theta}_R)$ \\
			$W^- \mu^+ \nu$ & $\displaystyle \frac{g_W}{2\sqrt{2}}\cos{\theta}_L$ & $\displaystyle \frac{g_W}{2\sqrt{2}}\cos{\theta}_L$ \\
            \hline
		\end{tabular}
  }
		\caption{The first block presents the relevant Scalar Fermion Fermion (SFF) vertices in the form of $C_S-C_P\gamma_5$ where $C_S$ and $C_P$ are the scalar and psuedo-scalar couplings, respectively. The second block presents relevant Vector Fermion Fermion (VFF) vertices in the form of $\gamma^\mu\left(C_V-C_A\gamma_5\right)$ where $C_V$ and $C_A$ are the vector and axial vector parts, respectively. }
		\label{tab:couplings of VVA/VVZ vertex for MPVDM}
	\end{table}

\begin{figure}[htbp]
\centering
\includegraphics[width=0.5\textwidth]{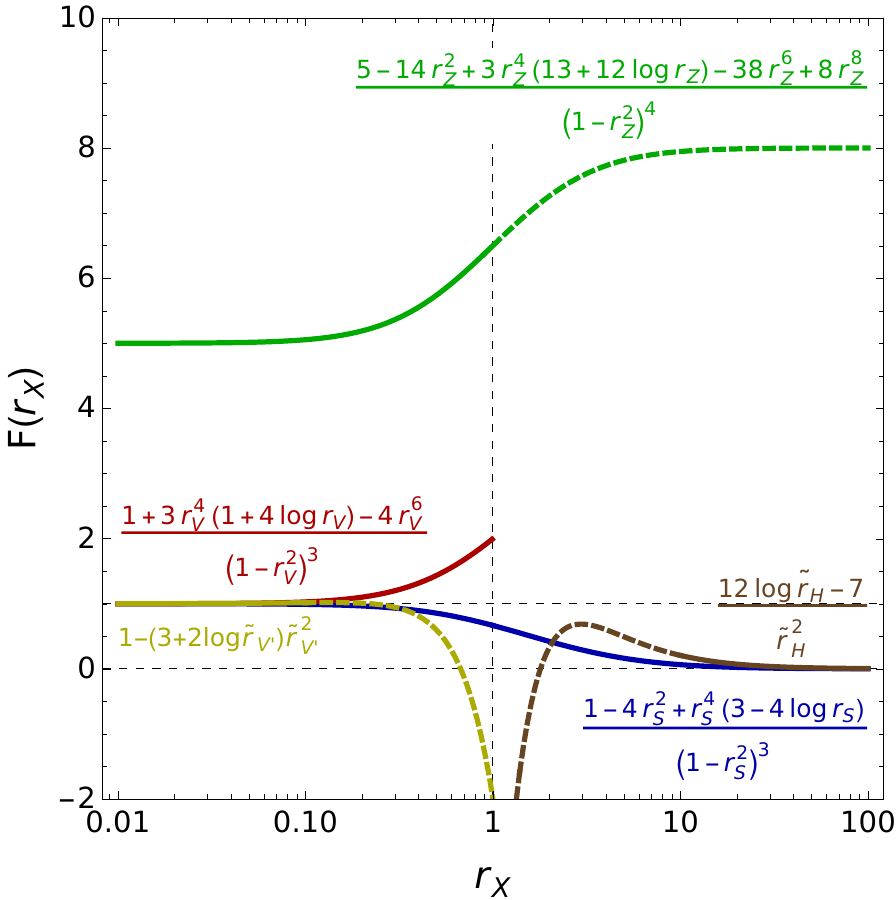}%
\caption{\label{fig:FrSV} Loop functions appearing in the scalar and vector contributions to $a_\mu$ in the MPVDM. The blue line corresponds to the contribution from scalar-$\mu'$ loop.  The brown line shows the loop function for scalar-$\mu$ loop. The red line corresponds to the 
$V^\prime\mu^\prime$ loop contribution, the green line to $Z\mu^\prime$, and the olive line to $V^\prime\mu$. 
When relations are valid only in specific limits (see text), the regions where the
limit is not achieved are represented by dashed lines.
The range of $r_{V_D}$ and $r_{V^\prime}$ is strictly bounded between 0 and 1 due to the model hierarchy: $m_{V^\prime} < m_{V_D} < m_{\mu_D} < m_{\mu^\prime}$.}
\end{figure}

If is convenient to define the following mass ratios:
\begin{eqnarray}
\begin{array}{ccc}
r_\mu={m_\mu\over m_{\mu_D}}\;,~&
r_D={m_{\mu_D}\over m_{\mu^\prime}}\;,~&
r_F=r_\mu r_D={m_{\mu}\over m_{\mu^\prime}}\;,~
\end{array}\\
\begin{array}{cccc}
r_H={m_H\over m_{\mu^\prime}}\;,~&
\tilde r_H={m_H\over m_\mu}\;,~&
r_{H_D}={m_{H_D}\over m_{\mu^\prime}}\;,~&
\tilde r_{H_D}={m_{H_D}\over m_\mu}\;,~
\end{array}
\end{eqnarray}
With these definitions, and in the limit of small $r_\mu$ and thus small $r_F$, the contributions from $a_\mu^{(a)}$ read:
\begin{align}
a_\mu^{(a)}[H\mu^\prime] &= {g_W^2 m_\mu^2 \over 64\pi^2 m_W^2} \times r_\mu^2(1-r_D^2) \times {1-4r_H^2+r_H^4(3-4 \log r_H) \over (1-r_H^2)^3}\;,
\label{eq:amu-hmup}
\\
a_\mu^{(a)}[H_D\mu^\prime] &= -{g_D^2 m_\mu^2 \over 64\pi^2 m_{V_D}^2} \times (1-r_D^2)^2 \times {1-4r_{H_D}^2+r_{H_D}^4(3-4\log r_{H_D}) \over (1-r_{H_D}^2)^3}\;.
\label{eq:amu-hdmup}
\end{align}
Both contributions in ~\Cref{eq:amu-hmup,eq:amu-hdmup} can be factorised into three
components. The first is proportional to $m_\mu^2$ over the square of the gauge boson mass associated
with the scalar's gauge group (either $m_W$ or $m_{V_D}$), multiplied by the respective gauge
coupling. This term is always positive and sets the overall scale of the contribution. The second
factor, which depends on fermion mass ratios, is also always positive. Crucially, this factor is
significantly larger for $H_D$ than for $H$, since the $H$ contribution carries an additional
suppression by $r_\mu^2 \ll 1$. Therefore, the dark Higgs loop typically dominates in magnitude -
although it contributes with a negative sign due to the overall minus in \Cref{eq:amu-hdmup}. The
third factor is a loop function of the scalar-to-fermion mass ratio; it is always positive and
decreases rapidly with increasing scalar mass, as shown by the blue line in~\Cref{fig:FrSV}.

Moving to the $a_\mu^{(b)}$ contributions, still in the limit of small $r_\mu$, we obtain:
\begin{eqnarray} 
a_\mu^{(b)}[H\mu]_{\rm NP} &=& \frac{1}{3} \frac{g_W^2 m_\mu^2}{64\pi^2 m_W^2}
\times \frac{(1 - r_D)^2}{r_D^2} \times \frac{12 \log \tilde r_H - 7}{\tilde r_H^2}\;, \\
a_\mu^{(b)}[H_D\mu]_{m_\mu \ll m_{H_D}} &=& \frac{1}{3} \frac{g_D^2 m_\mu^2}{64\pi^2 m_{V_D}^2}
\times (1 - r_D^2)^2 \times \frac{12 \log \tilde r_{H_D} - 7}{\tilde r_{H_D}^2}\;, \\
a_\mu^{(b)}[H_D\mu]_{m_{H_D} \ll m_\mu} &=& 3 \frac{g_D^2 m_\mu^2}{64\pi^2 m_{V_D}^2} \times (1 -
r_D^2)^2\;, 
\end{eqnarray}
where in $a_\mu^{(b)}[H\mu]_{\rm NP}$ we have subtracted the pure SM contribution to isolate the new physics terms.
These contributions differ by their numerical prefactors, which can enhance or suppress them relative
to one another. The first factors in all three expressions are structurally analogous to the
$a_\mu^{(a)}$ case. 
The ratio between the $H$ and $H_D$ contributions in the second factor is
proportional to $1/\left(r_D^2(1 + r_D)^2\right)$. This ratio is large for small $r_D$, drops below
unity for $r_D > (\sqrt{5} - 1)/2$, and approaches $1/4$ in the decoupling limit $r_D = 1$. 
The third factor, involving the scalar-to-muon mass ratio, takes a definite numerical value in the
case of the $H$ contribution, approximately $5.56 \times 10^{-5}$. For the $H_D$ contribution, this
factor is always positive in the limit $m_{H_D} \gg m_\mu$, as shown by the brown line in~\Cref{fig:FrSV}.

The contribution from diagrams involving a neutral vector $V$ and a muon or muon partner $f$ is given by:
\begin{equation}
\small
a_\mu^V = \frac{m_\mu^2}{4\pi} \int_0^1 dx \, \frac{C_V^2\left[x(1-x)\left(x - 2 + \frac{2m_f}{m_\mu}\right) - \frac{m_\mu^2}{2m_V^2} \left(x^3\left(\frac{m_f}{m_\mu} - 1\right)^2 + x^2\left(\frac{m_f^2}{m_\mu^2} - 1\right)\left(1 - \frac{m_f}{m_\mu}\right)\right)\right] + C_A^2(m_f \to -m_f)}{m_\mu^2 x^2 + (m_f^2 - m_\mu^2)x + m_V^2(1 - x)}\;,
\end{equation}
where $C_V$ and $C_A$ are the vector and axial-vector couplings in the vector–fermion–fermion vertices, and $m_V$ is the mass of the gauge boson circulating in the loop ($Z$, $V^\prime$, or $V_D$).

If $f = \mu^\prime$ or $\mu_D$ and $V = V^\prime$ or $V_D$ (\Cref{fig:feyngm2c,fig:feyngm2d}), the integration gives:
\begin{eqnarray}
a_\mu^{(c,d)}[Vf] &=& \frac{(C_V^2 - C_A^2)}{16\pi^2} \frac{m_\mu m_f}{m_V^2 (m_f^2 - m_V^2)^3} \left( m_f^6 + 3m_f^2 m_V^4 \left(1 - 4\log\left(\frac{m_f}{m_V}\right)\right) - 4m_V^6 \right) \nonumber \\
&-& \frac{(C_V^2 + C_A^2)}{48\pi^2} \frac{m_\mu^2}{m_V^2 (m_f^2 - m_V^2)^4} \left( 5m_f^8 - 14m_f^6 m_V^2 + 3m_f^4 m_V^4 \left(13 - 12\log\left(\frac{m_f}{m_V}\right)\right) - 38 m_f^2 m_V^6 + 8 m_V^8 \right) \nonumber \\
&+& \mathcal{O}\left(\frac{m_\mu^4}{m_f^4}\right) + \mathcal{O}\left(\frac{m_V^2}{m_f^2}\right)\;,
\end{eqnarray}
where it is notable that the contribution proportional to $C_V^2 + C_A^2$ appears only at second order in the $m_\mu$ expansion.

When $f = \mu$ (\Cref{fig:feyngm2e}), the integration yields:
\begin{eqnarray}
m_\mu \ll m_V: &&\quad a_\mu^{(e)}[V\mu] = \frac{(C_V^2 - 5 C_A^2)}{48\pi^2} \frac{m_\mu^2}{m_V^2} + \mathcal{O}\left(\frac{m_\mu^4}{m_V^4}\right)\;, \\
m_V \ll m_\mu: &&\quad a_\mu^{(e)}[V\mu] = \frac{1}{8\pi^2} \left(C_V^2 + C_A^2\left(5 - 2\frac{m_\mu^2}{m_V^2} - 4\log\left(\frac{m_\mu}{m_V}\right)\right)\right) + \mathcal{O}\left(\frac{m_V^4}{m_\mu^4}\right)\;.
\label{eq: NC result}
\end{eqnarray}

We now define the following mass ratios relative to the vectors:
\begin{eqnarray}
\begin{array}{cccc}
r_Z = \frac{m_Z}{m_{\mu^\prime}}\;, \quad
r_{V^\prime} = \frac{m_{V^\prime}}{m_{\mu^\prime}}\;, \quad
r_{V_D} = \frac{m_{V_D}}{m_{\mu_D}}\;, \quad
\tilde r_{V^\prime} = \frac{m_{V^\prime}}{m_\mu}\;.
\end{array}
\end{eqnarray}

With these definitions, and substituting the relevant couplings from \Cref{tab:couplings of VVA/VVZ vertex for MPVDM}, the $a_\mu^{(c)}$ contributions in the small $r_\mu$ limit are:
\begin{eqnarray}
a_\mu^{(c)}[Z\mu^\prime] &=& -\frac{1}{6} \frac{g_W^2 m_\mu^2}{64\pi^2 m_W^2} \times r_\mu^2 (1 - r_D^2) \times \frac{5 - 14r_Z^2 + 3r_Z^4(13 + 12\log r_Z) - 38r_Z^6 + 8r_Z^8}{(1 - r_Z^2)^4}\;, \\
a_\mu^{(c)}[V^\prime\mu^\prime] &=& \frac{g_D^2 m_\mu^2}{64\pi^2 m_{V^\prime}^2} \times (1 - r_D^2) \times \frac{1 + 3r_{V^\prime}^4(1 + 4\log r_{V^\prime}) - 4r_{V^\prime}^6}{(1 - r_{V^\prime}^2)^3}\;.
\end{eqnarray}

The structure of these expressions mirrors that of the scalar contributions. The first factor is analogous, while the $Z$ loop is further suppressed by $r_\mu^2$ compared to the $V^\prime$ loop. This suppression reflects the fact that the leading $Z$ contribution arises only at second order in $m_\mu$, being proportional to $C_V^2 + C_A^2$.
The loop functions in the last factors depend on $r_Z$ and $r_{V^\prime}$. These functions are shown in~\Cref{fig:FrSV}, where the $V^\prime$ case is represented by the red line, while $Z$ case is presented by the green line. For the $Z$ loop, only the region with $m_{\mu^\prime} > m_Z$ is relevant, due to collider bounds on $\mathbb{Z}_2$-even muon partner masses, discussed in~\Cref{sec:collider pheno}. Both functions are strictly positive, making the $Z$ contribution negative and the $V^\prime$ contribution positive. Moreover, the loop functions are always larger than one, leading to an overall enhancement of the vector contributions.

Moving to the $a_\mu^{(d)}$ contribution, we obtain:
\begin{eqnarray}
a_\mu^{(d)}[V_D\mu_D] &=& \frac{1}{2} \frac{g_D^2 m_\mu^2}{64\pi^2 m_{V_D}^2} \times (1 - r_D^2) \times \frac{1 + 3r_{V_D}^4(1 + 4\log r_{V_D}) - 4 r_{V_D}^6}{(1 - r_{V_D}^2)^3}\;.
\end{eqnarray}
This contribution is always positive and becomes bigger
with $m_{V_D}$ decrease and/or with $r_{V_D}$
increase,
see the red line in~\Cref{fig:FrSV}.

The next set of contributions comes from $a_\mu^{(e)}$:
\begin{eqnarray}
a_\mu^{(e)}[Z\mu]_{\rm NP} &=& -\frac{4}{3} \frac{g_W^2 m_\mu^2}{64\pi^2 m_W^2} \times r_\mu^4 (1 - r_D^2)^2\;, \\
a_\mu^{(e)}[V^\prime\mu]_{m_\mu \ll m_{V^\prime}} &=& -\frac{4}{3} \frac{g_D^2 m_\mu^2}{64\pi^2 m_{V^\prime}^2} \times (1 - r_D^2)^2\;, \\
a_\mu^{(e)}[V^\prime\mu]_{m_{V^\prime} \ll m_\mu} &=& -\frac{g_D^2 m_\mu^2}{64\pi^2 m_{V^\prime}^2} \times (1 - r_D^2)^2 \times \left(1 - (3 + 2\log \tilde r_{V^\prime}) \tilde r_{V^\prime}^2\right)\;, \label{eq:amuesmallVp}
\end{eqnarray}
where in the $Z$ loop we have subtracted the SM contribution. These contributions are always negative. The behaviour of the third term in Eq.~\eqref{eq:amuesmallVp} 
from $V'$ loop is shown by the olive line in~\Cref{fig:FrSV}. The $Z$ loop is further suppressed by a factor of $r_\mu^4$ compared to the $V^\prime$ loop.

Finally, the contribution from the charged current in $a_\mu^{(f)}$, involving $W$ and a neutrino, is given by:\footnote{The full expression for the loop with a generic charged vector and neutral fermion is given in \cite{Leveille:1977rc}. We do not reproduce it here since there are no new charged vectors or heavy neutral fermions in the MPVDM.}
\begin{eqnarray}
a_\mu^W &=& \frac{C_V^2 + C_A^2}{4\pi^2} \times \frac{5}{6} \frac{m_\mu^2}{m_W^2}\;,
\end{eqnarray}
which, after substituting the couplings and subtracting the SM piece, becomes:
\begin{eqnarray}
a_\mu^{(f)}[W\nu]_{\rm NP} &=& -\frac{5}{6} \frac{g_W^2 m_\mu^2}{64\pi^2 m_W^2} \times r_\mu^4 (1 - r_D^2)^2\;,
\end{eqnarray}
a result that is extremely small, even if the two muon partners are nearly degenerate.

\vspace{0.5em}
The above expressions provide analytical insight into new physics contributions across wide regions of the parameter space. In the next section, we present numerical evaluations based on full integration of the loop expressions to accurately capture intermediate regimes and subleading effects.

\bibliographystyle{JHEP}
\bibliography{ref}

\end{document}